\RequirePackage{fix-cm}
\documentclass[smallextended]{svjour3}       
\smartqed  
\usepackage{booktabs} 
\usepackage{amsmath,amssymb,amsfonts}
\usepackage{algorithm}
\usepackage{algorithmic}
\usepackage{graphicx}
\usepackage{textcomp}
\usepackage{xcolor}
\usepackage{multirow}
\usepackage{subfigure}
\usepackage[super]{nth}
\usepackage{natbib}
\setcitestyle{aysep={}} 

\graphicspath{ {imgs/} }

\newcommand{\wenyi}[1]{{\color{black}{#1}}}

\def\bW{\mathbf{W}}
\def\bA{\mathbf{A}}

\def\bK{\mathbf{K}}

\def\bw{\mathbf{w}}
\def\bb{\mathbf{b}}
\def\be{\mathbf{e}}
\def\ba{\mathbf{a}}
\def\bd{\mathbf{d}}
\def\bh{\mathbf{h}}

\def\be{\mathbf{e}}
\def\bf{\mathbf{f}}
\def\bu{\mathbf{u}}
\def\bq{\mathbf{q}}
\def\bo{\mathbf{o}}
\def\bx{\mathbf{x}}
\def\by{\mathbf{y}}
\def\bh{\mathbf{h}}
\def\bv{\mathbf{v}}
\def\bs{\mathbf{s}}
\def\bu{\mathbf{u}}

\DeclareMathOperator*{\argmax}{argmax}

\begin{document}

\title{Social Explorative Attention based Recommendation for Content Distribution Platforms
}

\titlerunning{SEAN}        

\author{Wenyi Xiao         \and
        Huan Zhao \and
        Haojie Pan \and
        Yangqiu Song \and
        Vincent W. Zheng \and
        Qiang Yang 
}

\authorrunning{Social Explorative Attention Network} 

\institute{Wenyi Xiao \at
              HKUST, Hong Kong, China \\
            \email{wxiaoae@cse.ust.hk}           
           \and
           Huan Zhao \at
              4Paradigm Inc., China\\
             \email{zhaohuan@4paradigm.com}
           \and
           Haojie Pan \at
              HKUST, Hong Kong, China \\
              \email{hpanad@cse.ust.hk} 
            \and
            Yangqiu Song \at
              HKUST, Hong Kong, China \\
              Peng Cheng Laboratory, Shenzhen, China\\
              \email{yqsong@cse.ust.hk} 
            \and
            Vincent W. Zheng \at
              WeBank, China \\
              \email{vincentz@webank.com} 
            \and
            Qiang Yang \at
              WeBank, China \\
              \email{qiangyang@webank.com}
}

\date{Received: 23 February 2020 / Accepted: 2 December 2020}

\maketitle

\begin{abstract}
In modern social media platforms, an effective content recommendation should benefit both creators to bring genuine benefits to them and consumers to help them get really interesting content.
To address the limitations of existing methods for social recommendation, we propose Social Explorative Attention Network (SEAN), a social recommendation framework that uses a personalized content recommendation model to encourage personal interests driven recommendation. 
SEAN has two versions: (1) SEAN-END2END allows user's attention vector to attend their personalized interested points in the documents.
(2) SEAN-KEYWORD extracts keywords from users' historical readings to capture their long-term interests. It is much faster than the first version, more suitable for practical usage, while SEAN-END2END is more effective.
Both versions allow the personalization factors to attend to users' higher-order friends on the social network to improve the accuracy and diversity of recommendation results. 
Constructing two datasets in two languages, English and Spanish, from a popular decentralized content distribution platform, Steemit, we compare SEAN models with state-of-the-art collaborative filtering (CF) and content based recommendation approaches. 
Experimental results demonstrate the effectiveness of SEAN in terms of both Gini coefficients for recommendation equality and F1 scores for recommendation accuracy.

\end{abstract}

\section{Introduction}
\label{sec-intro}
With the prevalence of online social platforms,  content recommendation has developed as a promising direction that leverages some side information such as social connections among users to enhance recommendation performance.
When applying to modern content distribution platforms, such as Facebook and Steemit\footnote{Steemit (https://steemit.com/) is a blockchain based social media and decentralized content distribution platform for consumers and creators to earn Steemit tokens by playing with the platform and interacting with others. It is regarded as a more effective content distribution ecosystem that allows small content creators to share their creative contents while protecting the copyright without any intermediaries. }, an effective content recommendation algorithm should consider both content creators to bring genuine benefits to them and content consumers to help them get really interesting contents.
While more accurate recommendation can improve the consumers' reading experience, it is regarded as a healthier content distribution ecosystem that encourages individuals to create contents. However, existing recommendation algorithms still lack consideration for balancing both content creators and consumers.

In the literature, content recommendation methods can be content based or collaborative filtering (CF) based ones. Content based methods~\citep{wang2017dynamic,wang2018dkn} memorize historical reading/watching content of a user and predict his/her future reading/watching content based on features or similarities of both contents. 
Such approaches emphasize particular topics for a user and may not be able to encourage diversity of recommendation results unless a content consumer actively searches for new topics. On the other hand, CF is considered as a complementary technique for content recommendation~\citep{das2007google} as it recommends based on similar users' clicking behaviors in the whole platform.
CF usually optimizes based on global behavior information so that the platform will attract more clicks or reading actions.
Despite the recommending effectiveness, CF will produce unintended Matthew's Effects (``The Rich Get Richer'')~\citep{MatthewEffect} which
will hurt small/new content creators who may not be able to attract attentions.
Although most of the traditional CF based methods are often called personalized recommendation~\citep{das2007google,liu2010personalized} and can be generalized to social networks using social regularization~\citep{jamali2010matrix,ma2011recommender,ye2012exploring,yang2012circle,zhao2017sloma,sun2018attentive,chen2019social}, they are in nature looking at global information and cannot solve this problem.

\begin{figure}[t]
\centering
\includegraphics[width=0.7\textwidth]{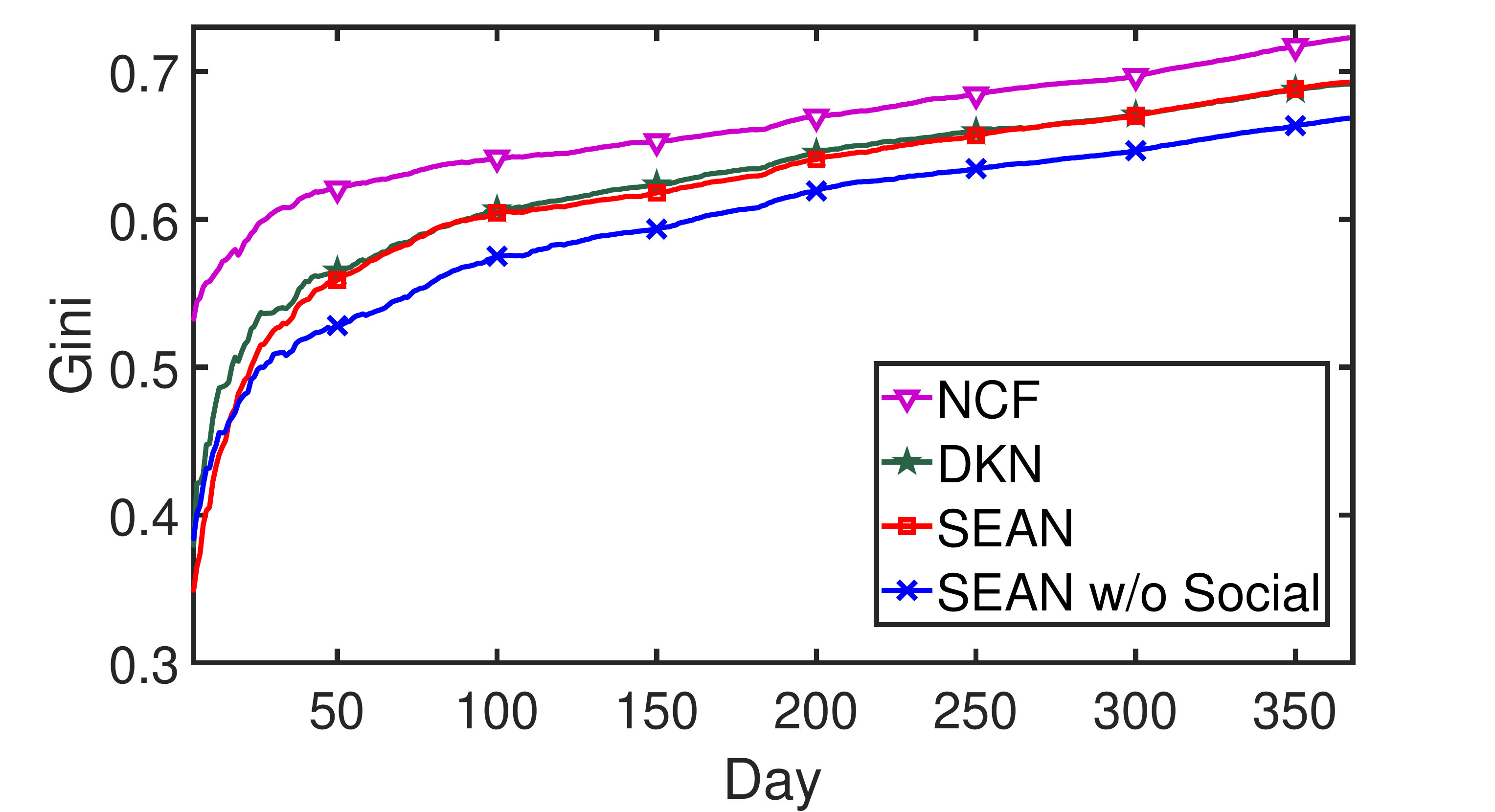}
\caption{Comparison of Gini coefficients of different algorithms for 368 days using Steemit social media. 
Gini coefficient is computed over the distribution of recommendation impression numbers of content creators.
We compared content based DKN~\citep{wang2018dkn} (F1=42.85), NCF~\citep{he2017neural} (F1=42.14), and our algorithm SEAN with (F1=47.69) and without any social information (F1=42.40). 
}
\label{gini-comparison}
\end{figure}

A few RSs ~\citep{SalganikDodds,AbeliukBHHL17} analyze the Matthew's Effect.
However, they are still CF based recommendation and the recommendation strategies are relatively simple, e.g., using popularity~\citep{SalganikDodds} or quality~\citep{AbeliukBHHL17,BerbegliaH17} of the content.
Moreover, one possible way to reduce Matthew's Effect is to use mechanism design approaches in game theory~\citep{BerbegliaH17}. 
In fact, the developed strategy (e.g., introducing randomness in recommendation~\citep{BerbegliaH17}) may hurt consumer's satisfaction as the recommended content may not be related to a consumer's interests, but such effect has not been considered and discussed yet.
As shown in Figure~\ref{gini-comparison}, we use the Gini coefficient over the distribution of recommendation impression numbers for content creators on Steemit social media to demonstrate Matthew's Effect, as the Gini coefficient is usually used to measure inequality and large Gini values mean eventually a small number of content creators dominating the content consumption. 
We also use the F1 score of prediction to evaluate content consumer's satisfaction.
From the Figure~\ref{gini-comparison}, we can see that the state-of-the-art CF based approach NCF~\citep{he2017neural}  encourages inequality more than content based approach DKN~\citep{wang2018dkn} although their F1 scores are comparable to each other.
Thus, a natural question is {\it can we have a content recommendation algorithm that can further benefit content consumers while not hurting creators?}

In this article, we propose the social explorative attention network (SEAN) to consider both creators and consumers.
For {\it creator equality}, we use a content based approach.
However, as a traditional content based approach, DKN encodes the new incoming document into a unique vector (the same vector that will be compared with all users) and uses this vector to attend to a user's historically read documents. 
In this way, popular content will still tend to be selected by the final prediction classifier regardless of the user's interests.
Different from DKN, we use user-dependent vectors to attend to related words and sentences in a new incoming document.
In this way, we compress a user's interests into contextual vectors and use the user-dependent document representation vector to feed the final prediction classifier.
This is more compatible with the personalized nature of content recommendation that can benefit small creators, as long as the created content is of the consumer's interests.
As shown in Figure~\ref{gini-comparison}, our model without social information can achieve comparable F1 while significantly reduce the Gini coefficient.
A natural way to further encourage diversity and improve creators' equality is to introduce more randomness as suggested by the mechanism design approach~\citep{BerbegliaH17}. We demonstrate this by randomly exploring other consumers' interests. However, this mechanism will hurt prediction accuracy which reflects the consumer's satisfaction.

To improve the {\it consumer satisfaction}, as the training data for each user's context vectors may be too small and cannot train well for users having less reading history, we design two variants of SEAN, SEAN-END2END and SEAN-KEYWORD, to address it. In SEAN-END2END, we allow a user's context vectors to ``attend'' to friends' context vectors and fetch back friends' reading interests and prediction knowledge by aggregating their context vectors. In SEAN-KEYWORD, we extract keywords and use a word-level attention module to attend to the user's interested words in both documents and users' historical readings.
Nonetheless, even a user has many one-hop friends, friends sharing a similar topic of interest may not be enough.
Therefore, we consider the user's higher-order friends.
An extreme case is that we go over all $n$-hop friends, and we can likely reach all connected users in a social network when $n$ is large.
Apparently, it will be too expensive to explore.
To remedy this, we develop a social exploration mechanism based on Monte Carlo Tree Search (MCTS)~\citep{MasteringGo2016}.
This will be more effective to attend to higher-order friends.
In particular, by using MCTS, we can achieve a good balance of finding $n$-hop friends with similar interests and exploring friends with some randomness for more diverse interests in a social network.
As shown in Figure~\ref{gini-comparison}, introducing social information will significantly improve the F1 score but also increase the Gini coefficient.
Therefore, in the experiments, we systematically study how different hyper-parameters of SEAN, including social information, can affect both prediction accuracy and the Gini coefficient.

Our contributions can be highlighted as follows:
\begin{itemize}
    \item 
    To the best of our knowledge, this is the first study on comprehensive exploring of news recommendation with social connections on decentralized platforms, where the creator equality is much more important than traditional content distribution platforms. In particular, we use the Gini coefficient to measure the inequality of content creators based on recommendation impressions.
    
    \item We propose a novel social explorative attention based recommendation model, SEAN, to use a user's personal reading history and go beyond personal data to explore the user's higher-order friends. We compare two variants, SEAN-END2END and SEAN-KEYWORD, by considering both effectiveness and efficiency.
    
    \item We construct two datasets of different languages from a popular decentralized content distribution platform, Steemit. By conducting extensive experiments, we demonstrate the superiority of our model over existing state-of-the-art recommendation approaches, including CF and content based ones, in terms of benefiting both consumers and creators.
\end{itemize}

\wenyi{Preliminary results of this manuscript have been reported in \citep{xiao2019sean}, published as a conference paper in SIGKDD 2019,
where MCTS is designed to explore high-order friends for the target user, and then a personalized hierarchical attention network is proposed for news recommendation. 
In this full version, we propose and highlight the content modeling and friend selection for the user, which provides a comprehensive solution for fusing social information for RS on decentralized platforms.
We conduct extensive research on social exploration strategy designing, content modeling, which constitutes a novel and effective user-item interaction model for more industry-driven applications. Hence, the contributions of this work lie in broader domains. To be specific, for social exploration, besides MCTS, we propose to use $\epsilon$-Greedy as another strategy to balance exploitation and exploration, introduced in Section~\ref{sec-greedy-exploration}.
For content modeling, we propose two versions of SEAN, which have different interactions among documents and users: 
(1) The end-to-end version (SEAN-END2END) uses user-dependent vectors to attend to related words and sentences in a new incoming document, introduced in Section~\ref{sec-sean-end2end}. This is the original model(SEAN) introduced in \citep{xiao2019sean}.
(2) The keyword version (SEAN-KEYWORD), introduced in Section~\ref{sec-sean-keyword}, first extracts keywords, and then a use word-level attention module to attend to user's interested words in both newly incoming documents and users' historical readings dynamically. Furthermore, for incorporating social information, we propose both dynamic attention and static attention (various similarity functions) for weighing the influence of users' friends. 
SEAN-END2END is more effective while SEAN-KEYWORD runs much faster.
Additional experiments are performed, to support the increased components in Section~\ref{sec-vs-baselines}, \ref{sec-vs-exploration}, \ref{sec-two-sean-comparison}, \ref{sec-sean-variant} and \ref{sec-param-sens}, respectively.
Finally, we give a comprehensive review on related works in Section \ref{sec-related} and point out some potential research for the future works in Section \ref{sec-conclusion}.}

We release the code and datasets for researchers to validate the reported results and conduct further research. The code and data are available at https://github.com/HKUST-KnowComp/Social-Explorative-Attention-Networks.


\section{Overview}
In this section, we first give the descriptions of the problem we plan to solve and list the key notations used in the article. Then we give a brief introduction of model frameworks.

\begin{table}[]
\centering
\caption{Key notations in this article.}
\label{tb-note}
{
\begin{tabular}{cc}
\toprule
\textbf{Notation} & \textbf{Meaning} \\
\midrule
$\mathcal{G} = (\mathcal{U}, \mathcal{E})$ & social graph $\mathcal{G}$, set of users $\mathcal{U}$, set of edges $\mathcal{E}$ \\
$Q_t(v)$ & exploitation reward of node $v$ at day $t$ \\
$U_t(v)$ & exploration score of node $v$ at day $t$ \\
$B$ & beam width \\
$L$ & search depth / \# of selected friends \\
$\lambda$ & trade-off constant \\
$\epsilon$ & probability to take a random action in $\epsilon$-Greedy \\
$\alpha$ & PageRank value for SPR \& DPR \\
$K$ & number of kernels in CNN \\
$g$ & window size in CNN \\
$r$ & filter size in CNN \\
$h$ & user embedding size \& hidden size \\
$D$ & the dimension of word embedding \\
$m$ & \# of keywords to represent user \\
$n$ & \# of keywords to represent document \\
\bottomrule
\end{tabular}
}
\end{table}

\subsection{Problem Formulation}
\wenyi{The recommendation task is to predict whether a target user $u$ will click a given document $d$. 
Here we use the textual document recommendation as an example for content recommendation.
We assume that we have a social graph $\mathcal{G} = (\mathcal{U}, \mathcal{E})$, where $\mathcal{U}$ is the set of users, and $\mathcal{E}$ is the set of edges, representing the social connection between two users.
Our goal is to learn a prediction function $p = \mathcal{F}\left ( u, d, \theta  \right )$, where $p$ represents the probability that user $u$ will click a given document $d$, and $\theta$ denotes the model parameters of function $\mathcal{F}$. For the sake of clear presentation, we list the key notations used in this article in Table \ref{tb-note}}.

\subsection{Overall Framework}
In this article, we propose to use a personalized model to perform content recommendation, as personalization will encourage the model to find more relevant contents and less affected by the global information about popularity and social influence.
Then we socialize it to make the personalized factors be able to attend to friends' information, which will further balance the randomness factor to improve the creator equality and the relatedness factor to improve the consumer satisfaction.
To enable the attention over attention mechanism to use more information, we propose to explore a user's higher-order friends. 
We call our recommendation model  Social Explorative Attention Network (SEAN). \wenyi{And the details of two verions of SEAN, SEAN-END2END and SEAN-KEYWORD, will be introduced in the following subsection.}

For the whole procedure, we first initialize paths for users by randomly selecting users in the social graph and set users' explored times as the times they selected as friends. For each user in the $t$-th day training, we first explore $B$ sets of friends by MCTS strategy and update the explored times for these selected friends, consequently updating the exploration values $U_t(v)$ of them. 
These $B$ sets of friends are incorporated with the user as input to the RS model. Then we update $Q_t(u)$ of the target user. The updated results are used for the $t+1$-th day training. We show the algorithm for the model training in Algorithm~\ref{alg-sean}. \wenyi{Besides MCTS, we also propose to use $\epsilon$-Greedy as another strategy to balance exploitation and exploration and select high-order friends.}

\begin{algorithm}[htb]
    \caption{SEAN.}
    \label{alg-sean}
    \begin{algorithmic}[1]
        \FOR{$t = 1,2,\cdots$}
         \FOR{$u \in \mathcal{U}$}
             \STATE{${\{\mathcal{F}_i(u)}\}^B_{b=1}$ = SelectFriends(u, B, L)}
             \FOR{$b=1,2,\cdots,B$}
             \STATE{Train SEAN with $\mathcal{F}_b(u)$ for $u$;}
             \FOR{$v \in \mathcal{F}_b(u)$}
                 \STATE{$N_t(v) \leftarrow N_t(v) + \frac{1}{B}$;}
             \STATE{Update $U_t(v)$ according to Eq.~\eqref{eq-ucb-u};}
             \ENDFOR
                \STATE{Update $Q_t(u)$ for user $u$ according to Section ~\ref{sec-mcts-exploitation};}
            \ENDFOR
            \ENDFOR
        \ENDFOR
    \end{algorithmic}
\end{algorithm}


In the remaining part of this paper, we first introduce the social exploration in Section \ref{sec-social-exploration}, and then the two content modeling models in Section \ref{sec-sean-end2end} and Section \ref{sec-sean-keyword}, respectively. Further in Section~\ref{sec-exp} extensive experiments are conducted to demonstrate the effectiveness of SEAN in terms of recommendation equality and accuracy. Finally we review related work in Section~\ref{sec-related} and conclude this work in Section~\ref{sec-conclusion}.

\section{Friend Selection}
\label{sec-social-exploration}
In this section, we introduce two approaches to select relative high-order friends to enhance the effectiveness of our recommender system.  In section \ref{sec-ucb}, we demonstrate how we use MCTS for friends selection, and then further enhance MCTS with beam search. In section \ref{sec-greedy-exploration}, $\epsilon$-Greedy is introduced as another strategy to select high-order friends comparing to MCTS.

\subsection{Selecting Friends with MCTS}
\label{sec-ucb}

Monte Carlo Tree Search (MCTS)~\citep{MasteringGo2016} is a stochastic search algorithm to find an optimal solution in the decision space.
It models an agent that simultaneously attempts to acquire new knowledge (called ``exploration'') and optimize the decisions based on existing knowledge (called ``exploitation''). 
MCTS uses the upper confidence bounds one (UCB1)~\citep{kocsis2006bandit} value to determine the next move $a$ from a viewpoint of multi-armed bandit problem. The selection strategy is defined by: 
\begin{equation}
\label{eq-ucb1}
a = \argmax_v\{ Q_t(v) + \lambda \cdot U_t(v) \},
\end{equation}
where $Q_t(v)$ denotes the empirical mean exploitation reward of node $v$ at time $t$ and $U_t(v)$ is the utility to explore node $v$. This equation clearly expresses the exploration-exploitation trade-off: while the first term of the sum tends to exploit the seemingly optimal arm, the second term of the sum tends to explore less pulled arms. $\lambda$ is used to balance the two terms.

Here we explain how MCTS guides to generate a path with a fixed number of search depth $L$, regarded as $L$ friends of $u$ by walking through the social graph. We denote the target user $u$ as the starting node $c_0$ and denote $c_l, l\in \left [ 0, L \right ]$ as the $l$-th node added for $u$ in the path.
On day $t$ at search step $l$, the node $c_{l+1}$ is retrieved from the neighbors of node $c_l$. We calculate the score for each neighbor according to Eq.~\eqref{eq-ucb1} and choose the neighbor with the maximum score as the $(l+1)$-th friend of the user $u$.  The design of calculating the values from exploitation and exploration are mentioned below.

\subsubsection{Exploitation}
\label{sec-mcts-exploitation}
On day $t$, we compute the $Q_t(v)$ to get the exploitation reward of neighbor node $v$. In our scenarios,
we want to select those as friends who can improve the recommending performance as much as they can. In this work, we design four exploitation strategies to select friends for maximizing $Q_t(v)$: the average F1 from RS model (SEAN-RS-F1), static PageRank value from social network (SEAN-SPR), dynamic PageRank value from activity network (SEAN-DPR), as well as the actual payout each user earned in blockchain platforms (SEAN-Payout).
\begin{enumerate}
    \item \textbf{SEAN-RS-F1.} 
    We regard the average F1 evaluated based on our RS model of each neighbor node $v$ up to time $t$ as the exploitation reward $Q_t(v)$. This is based on the assumption that a user who has been well-learned by the RS model is reliable and could be exploited as a friend for the target user $u$ in the future. In this way, the RS prediction results can guide the friend exploration process, and in turn, the friend exploration process provides useful friends to help enrich the target user's representation. 
    \item \textbf{SEAN-SPR.}
    The second way is to use the PageRank value of $v$ obtained from social network as exploitation reward $Q_t(v)$. In \citep{xiang2013pagerank}, Xiang et.al. explicitly connect PageRank with social influence model and show that authority is equivalent to influence under their framework. Thus, we assume that a node with high PageRank value in the social network is influential and should be exploited as a friend for the target user $u$. 
    \item \textbf{SEAN-DPR.}
    On social media platforms, each user can not only make activities on the documents (as a consumer) but also create documents (as creator). We build a dynamic activity network and calculate the PageRank values of nodes. Compared with the social network, the edges in the activity network are the consumer-creator connection. 
    \item \textbf{SEAN-Payout.}
    In some blockchain based social platforms, the platform would give some rewards, i.e., bitcoin, to those users who help distribute the documents, i.e., post or forward a document in the platform. 
    We regard the payout that a user gains as the value of his/her exploitation value $Q_t(v)$.
\end{enumerate}

\begin{figure*}[]
\centering
\includegraphics[width=1\textwidth]{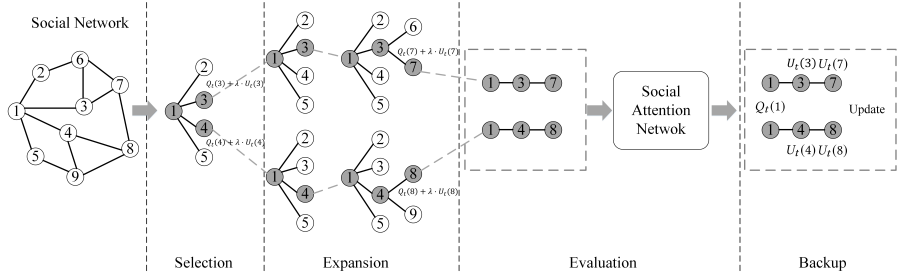}
\caption{MCTS for social exploration. We illustrate our MCTS based strategy with this example. We set the beam width and search depth to 2. The node '1' represents the target user. We initialize 2 paths according to the beam width and add "1" to each path. In the Selection step, we calculate the scores by Eq. (\ref{eq-ucb1}) of '1' 's neighbors ('2', '3', '4', '5') and select two nodes with the largest scores ('3', '4') and add them to each path. In the Expansion step,  we calculate the scores for the neighbors of both "3" and "4", and again select two nodes ('7', '8') among all the neighbors. In the Evaluation step, the two generated path ('1' $\rightarrow$ '3' $\rightarrow$ '7' $\rightarrow$   '1' $\rightarrow $ '4' $\rightarrow $ '8') are input to RS model. In the same way, we get the paths for other nodes. In the Backup step, we update $Q_t(1)$ from the result at Evaluation and $U_t(3), U_t(4), U_t(7), U_t(8)$ from Selection \& Expansion.}
\label{fig-mcts}
\end{figure*}

\subsubsection{Exploration}
\label{sec-mcts-exploration}
We design the exploration mechanism to get the explored reward $U_t(v)$ for friend $v$ as follows:
\begin{equation}
U_t(v) = \sqrt{\frac{\text{log} N_t(c_l)}{N_t(v)+ 1}},
\label{eq-ucb-u}
\end{equation}
where $N_t(c_l)$, $N_t(v)$ denote as the times that the current node $c_l$ at search step $l$ and the neighbor node $v$ have been selected as friends up to day $t$, respectively. The goal of the exploration is to select the nodes who have less been explored in the past.

\begin{algorithm}[htb]
    \caption{SelectFriends($u, B, L$).}
    \label{alg-bmcts}
    \begin{algorithmic}[1]
    \REQUIRE{target user $u$, beam width $B$, path length $L$}
         \ENSURE{$\big\{\mathcal{F}_b(u)\big\}_{b=1}^B$.}
     \STATE{\textbf{Initialization:} \\
     $\big\{\mathcal{F}_b(u)\big\}_{b=1}^B$: $\mathcal{F}_b(u)$ records the $b$-th path starting from $u$;\\
     $UCB1(v)$: UCB1 score for user $v$ according to Eq.~\eqref{eq-ucb1};\\
     $\big\{T_b(u)\big\}_{b=1}^B$: $T_b(u)$ records the sum of UCB1 scores of the $b$-th path for user $u$ during beam search;\\ 
    $\Delta_b$:the neighbours of the tail node of the path $\mathcal{F}_b(u)$ during beam search;
    }
    
    \WHILE{$k = 0,1,2,\cdots$, L:}
		\STATE{ $\mathcal{H}= \bigcup_{b=1}^B\big\{UCB1(v) + T_b(u), v\in \Delta_b\big\}$;}
		\WHILE{$b = 1,2,\cdots$, B:}
			\STATE{$v = \argmax_v\mathcal{H}$ ;}
			\STATE{$\mathcal{F}_b(u) \leftarrow \mathcal{F}_b(u) \bigcup v;$}
			\STATE{$\mathcal{H}\leftarrow \mathcal{H}\setminus (UCB1(v) + T_b(u))$;}
		\ENDWHILE
    \ENDWHILE
    \end{algorithmic}
\end{algorithm}

\subsubsection{Obtaining Multi-paths with Beam Search}
\label{sec-multi-mcts}
If we want to find higher-order friends, we can greedily select the next node with a maximum score from the neighbors of the current node at search step  $l$. In this way, we would get a path of higher-order friends. If we want to find more than one path, it is time-consuming to get a globally optimal set of paths. Therefore, we combine MCTS with beam search~\citep{koehn2004pharaoh} to balance the optimality and completeness. At search step $l$, we choose the neighbors with largest $B$ scores from Eq.~(\ref{eq-ucb1}) and these $B$ nodes are selected for further expansion. Here $B$ is the beam width. In this way, we generate $B$ paths for the target user $u$. For training and testing, we obtain $B$ prediction results by using each path and $u$ and compute the average of these results to get the final prediction. We give a concrete example of MCTS for social exploration, shown in Figure \ref{fig-mcts}. Besides, we give the whole procedure on how to select friends based on beam MCTS in Algorithm \ref{alg-bmcts}.

\subsection{Selecting Friends with $\epsilon$-Greedy}
\label{sec-greedy-exploration}
$\epsilon$-Greedy is the most popular and the simplest method to balance exploration and exploitation by taking the best action most of the time but do random exploration occasionally. We give a detailed introduction on how to incorporate this idea into the friend selection process in section \ref{sec-alg-greedy} and discuss the motivation and strength\&weakness of using $\epsilon$-Greedy in section \ref{sec-dis-greedy}.

\begin{algorithm}[ht]
    \caption{SelectFriends($u, B, L, \epsilon$).}
    \label{alg-greedy}
    \begin{algorithmic}[1]
    \REQUIRE{target user $u$, \# of path $B$, path length $L$, $\epsilon$}
         \ENSURE{$\big\{\mathcal{F}_b(u)\big\}_{b=1}^B$.}
     \STATE{\textbf{Initialization:} \\
     $\big\{\mathcal{F}_b(u)\big\}_{b=1}^B$: $\mathcal{F}_b(u)$ records the $b$-th path starting from $u$;\\
     $\Delta_b$:the neighbours of the tail node of the path $\mathcal{F}_b(u)$ during beam search;
    }
    \WHILE{$b = 1,2,\cdots$, B:}
        \WHILE{$k = 1,2,\cdots$, L:}
        \IF{$probability < \epsilon$:}
            \STATE{ $\mathcal{H}= \{Q(v)\}, v\in \Delta_b$;}
            \STATE{$v = \argmax_v\mathcal{H}$ ;}
        \ELSE
            \STATE{$v$ = Random $\Delta_b$;} 
        \ENDIF
        \STATE{$\mathcal{F}_b(u) \leftarrow \mathcal{F}_b(u) \bigcup v$;}
        \ENDWHILE
    \ENDWHILE
    \end{algorithmic}
\end{algorithm}

\subsubsection{Algorithm of $\epsilon$-Greedy}
\label{sec-alg-greedy}
In detail, for each step, with a probability $\epsilon$ we randomly select the friends without any bias. Meanwhile, with probability $1-\epsilon$ we select friends with higher exploitation value $Q_t(v)$ based on Section \ref{sec-mcts-exploitation}.
To obtain $B$ different paths, we follow the work in \citep{grover2016node2vec}, and the details are given in Algorithm \ref{alg-greedy}.

\subsubsection{Discussion on $\epsilon$-Greedy}
\label{sec-dis-greedy}
When settings $\epsilon$ to 0, it is a fully exploitative choice, thus the random walk process easily gets stuck to a finite set of vertices. It may not be good enough to explore the full graph. In practice, keeping a vaguely explorative/stochastic element in its policy (like a tiny amount of $\epsilon$) allows it to get out of such states. Compared to MCTS, $\epsilon$ is simple and straightforward to optimize.

Compared to MCTS, $\epsilon$ is simple and straightforward to optimize. Despite its simplicity, $\epsilon$-Greedy often yields pretty good results in some reinforcement learning tasks \citep{mnih2015human}. However, $\epsilon$-Greedy is less efficient to explore the most relative neighbors than MCTS since the exploration process in MCTS has supervising signals, consequently decreasing the searching time.

\section{SEAN-END2END}
\label{sec-sean-end2end}

\begin{figure}[]
\centering
\includegraphics[width=0.6\textwidth]{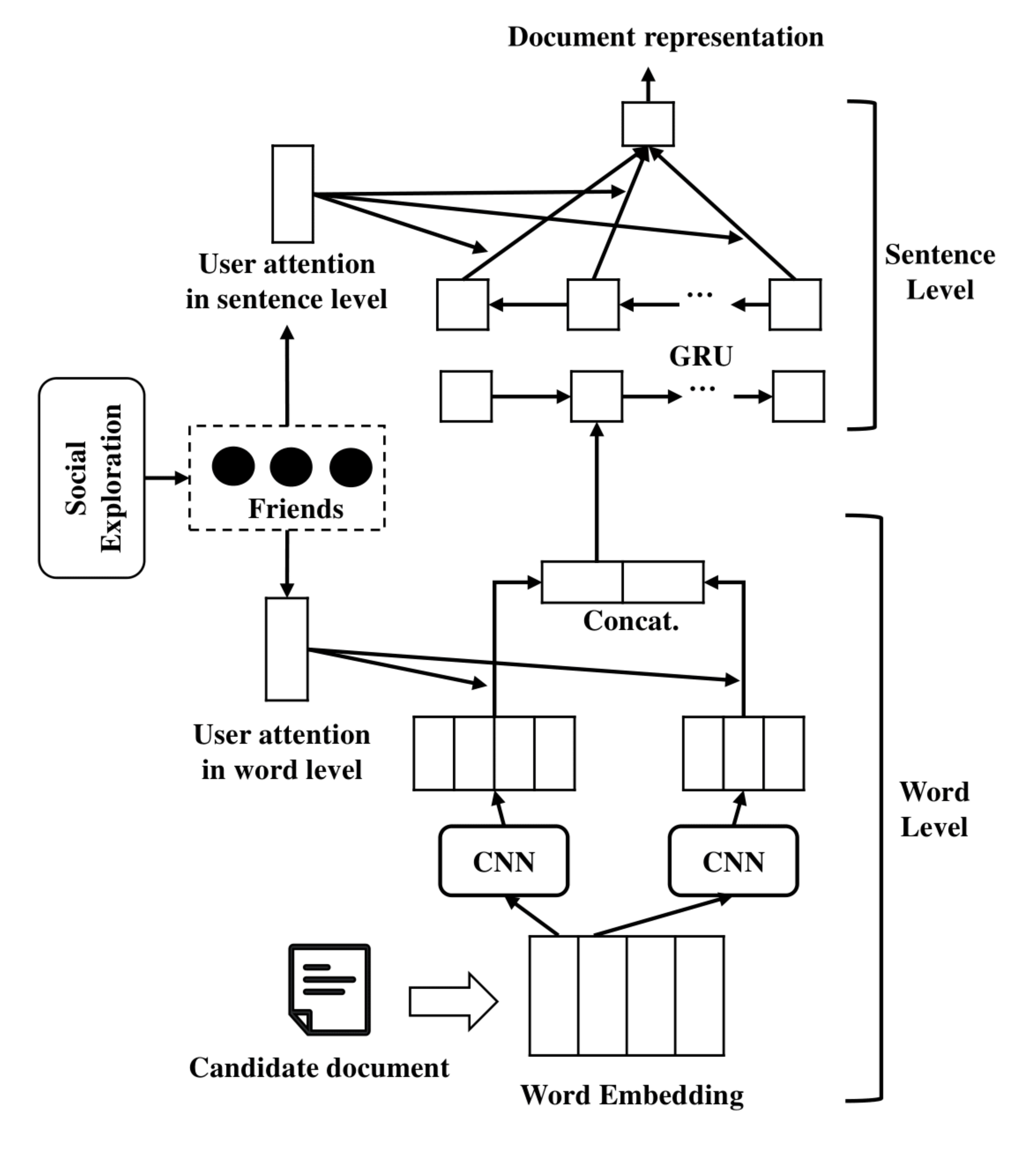}
\caption{The architecture of SEAN. The left side is a social exploration module that explores high-order friends for the RS system on the ride side. These friends are incorporated with the user to build the user's representation vector in word and sentence levels, respectively. The right side is a hierarchical architecture from CNN layer to encode words to GRU layer to encode sentences in the document. The user's representation vectors are used to attend to important words and sentences in the candidate document.}
\label{fig-sean-end2end}
\end{figure}

\wenyi{SEAN-END2END is an end-to-end framework to obtain better representation for users in social media with two modifications of recommendation: personalization and socialization. The overview architecture is shown in Figure~\ref{fig-sean-end2end}, and it contains three major components:
\begin{enumerate}
\item    Document Representation: we adopt the hierarchical attention networks~\citep{yang2016hierarchical}.
In the word level, the attention is used to select useful words to construct features for sentence representations.
\item    User Representation: we construct user representation vectors (word and sentence levels) themselves as attentions to his/her friends' representation vectors, which is essentially an attention over attention model.
\item    Output Layer: predicting the possibility that the candidate document will be clicked by the target user.
\end{enumerate}
}


\subsection{Socialized Document Representation}
\label{sec-single}
Assume that a document has $I$ sentences and each sentence contains $J$ words. $w_{ij}$ represents the $j$-th word in sentence $s_i$ with the indices $i \in \left [ 0, I\right ]$ and  $j\in \left [ 0, J \right ]$.
We use a hierarchical architecture to learn the document representation.

\subsubsection{Word Level Personalization}
We use pre-trained word embeddings for words and fix them during training. The embedding of each word can be calculated as $ \bw_{ij} = \bW_e \be_{w_{ij}}, i\in \left [ 0, I \right ], j\in \left [ 0, J \right ],
$
where $\bW_e$ is the embedding matrix for all words and $\be_{w_{ij}}$ is a one-hot vector to select one word embedding vector $\bw_{ij}$ for $w_{ij}$.
We concatenate all word embeddings in a sentence to form a sentence matrix $\bW_i \in \mathbb{R}^{D \times J}$ for sentence $s_i$, where $D$ is the dimension size of the embedding vector of each word.
We then use a convolutional neural network (CNN) \citep{kim2014convolutional} to represent sentences in the document. Here, we apply a convolution operation on $\bW_i$ with a kernel $\bK_k \in \mathbb{R}^{g\times r \times D}$, $k \in \left [ 0, K \right ]$ among $K$ kernels of width $g$ and filter size $r$ to obtain the feature $\bf^k$:
\begin{eqnarray}
\bf_{ij}^k = \text{relu}\left ( \bW_i\left [ \ast ,j:j+g-1 \right ]  \odot \bK_k + \bb_k \right),
\end{eqnarray}
where $j \in \left [ 1, J-g+1\right ]$ is the iteration index of convolution, $\bf_{ij}^k \in \mathbb{R}^{r}$ is regarded as $j$-th CNN feature by the $k$-th kernel $K_k$, and the bias vector $\bb_k$ in $i$-th sentence.

We then feed each CNN features $\bf_{ij}^k$ to a non-linear layer, parameterized by a global weight matrix $\bW_w \in \mathbb{R}^{{h}\times r}$ to get a hidden representation ${\bf_{ij}^k}'$. $h$ and $r$ are pre-defined dimensions of hidden vectors. 
We measure the importance of the word towards the target user as the similarity of ${\bf_{ij}^k}'$ and the word-level user's socialized representation vector $\bx_w$ (which will be introduced in Section \ref{sec-user}). The sentence representation vector $\bs_i^k$ by CNN with kernel size $k$ is computed as a weighted sum based on the soft attention weights:
\begin{eqnarray}
&& {\bf_{ij}^k}' =\tanh(\bW_w \bf_{ij}^k+\bb_w), \\
&& \alpha_{ij} = \text{Softmax}(\bx_w^{\top} {\bf_{ij}^k}'),\\
&& \bs_i^k=\sum_{j}\alpha_{ij} \bf_{ij}^k, 
\end{eqnarray}
where the superscript $\cdot^\top$ represents the vector or matrix transpose. All representation vector $\bs_i^k$ are concatenated together and taken as the sentence embedding $\bs_{i}$ for sentence $s_i$ as:
$
\bs_{i} = \left [ \bs_i^1, \bs_i^2, ..., \bs_i^K\right].
$

\subsubsection{Sentence Level Personalization}
At the sentence level, we use a bidirectional Gated Recurrent Unit network (Bi\-GRU) \citep{bahdanau2014neural} to compose a sequence of sentence vectors into a document vector. The BiGRU encodes the sentences from two directions:
\begin{equation}
    \bh_{i}  =   \overrightarrow{GRU} (\bs_{i}) \left |  \right |  \overleftarrow{GRU} (\bs_{i}).
\end{equation}
After getting $\bh_{i}$ for sentence $s_i$, we use the sentence level user representation vector $\bx_s$ to extract relevant sentences that are interested by the target user and get a final document representation $\bd$ by soft attention mechanism similar to sentence representation. We omit the details of equations due to the lack of space and the similarity with the word level computation.
As shown in the right side of Figure~\ref{fig-sean-end2end}, we have two layers of feature extraction networks.
This architecture is inspired by~\citep{yang2016hierarchical} since it is better for long document modeling.
In our model, we use CNN instead of RNN for word-level since in practice we found that CNN is faster, more robust, and less easy to overfitting on our datasets.
Moreover, different from \citep{yang2016hierarchical}, we use socialized user representation vectors instead of unified representation vectors for attending to words and sentences.

\subsection{Socialized User Representation}
\label{sec-user}
We denote $\be_u$ as a one-hot vector of user $u$ and retrieve the word level user representation $\bu_w$ from a trainable embedding matrix $\bA\in\mathbb{R}^{h\times |\mathcal{U}|}$ by using $\bA\be_u$, where $h$ is the size of attention vectors.
We can get the user's sentence-level representation by another trainable embedding matrix $\bA'$ in the same way.
We design a social attention module to enrich a user's representation by incorporating his/her friends' representations. 
Let $\by_i\in \mathbb{R}^{h}, i \in \left \{ 1,2,... \right \}$ be his/her friends' word-level representation vectors, and denote $\by_0 = u_w$. 
The attention mechanism produces a representation $\bx_w$ as a weighted sum of the representations vectors $ \by_j, j \in \left \{ 0,1,2,... \right \}$ via
\begin{eqnarray}
&& \alpha_j = \text{Softmax}(\text{LeakyReLU}(\bw^\top\left [ \bW_y \bu_w || \bW_y \by_j\right ])),\\
&& \bx_w = \sum_{j}\alpha_{j} \bW_y \by_j,
\end{eqnarray}
where $\bW_y \in \mathbb{R}^{h \times h}$ is a shared linear transformation and $||$ is the concatenation operation. The attention mechanism is a single-layer feedforward neural network, parametrized by a weight vector $\bw \in \mathbb{R}^{2h}$, and applying the LeakyReLU nonlinearity.

Similarly, we can get the sentence-level user representation $\bx_s$ by the attention of high-order friends' representation vectors.

\subsection{Prediction and Learning}

Finally, we use a dense layer to predict the probability that the target user $u$ will read the candidate document $d$:
\begin{equation}
p =\mathrm{Sigmoid} (\bw_g^\top \bd + \bb),
\end{equation}
where $\bw_g \in \mathbb{R}^{2h}$ is a global trainable weight vector trained by all the samples. $\bd$ is the document representation vector obtained from Section \ref{sec-single}.

Due to the nature of the implicit feedback and the task of item recommendation, we adopt the binary cross-entropy loss to train our model:
\begin{equation}
    \mathcal{L}(\theta)= -\frac{1}{M}\sum_{m=1}^{M} \left[y_m \log(p_m) + (1-y_m)\log(1-p_m) \right],
\end{equation}
where $m$ is the index of a sample, $M$ is the total number of training samples, $y_m\in \{0,1\}$ is the  label, and $\theta$ denotes the set of model parameters. The negative samples are formed from the documents that the target user does not make response to while his/her friends make. 

During training and testing, we train the model with the data of past $t$ days and test it with the data on $(t+1)$-th day. The model dynamically adapts to new data day-by-day.

\section{SEAN-KEYWORD}
\label{sec-sean-keyword}

\begin{figure}[]
\centering
\includegraphics[width=0.9\textwidth]{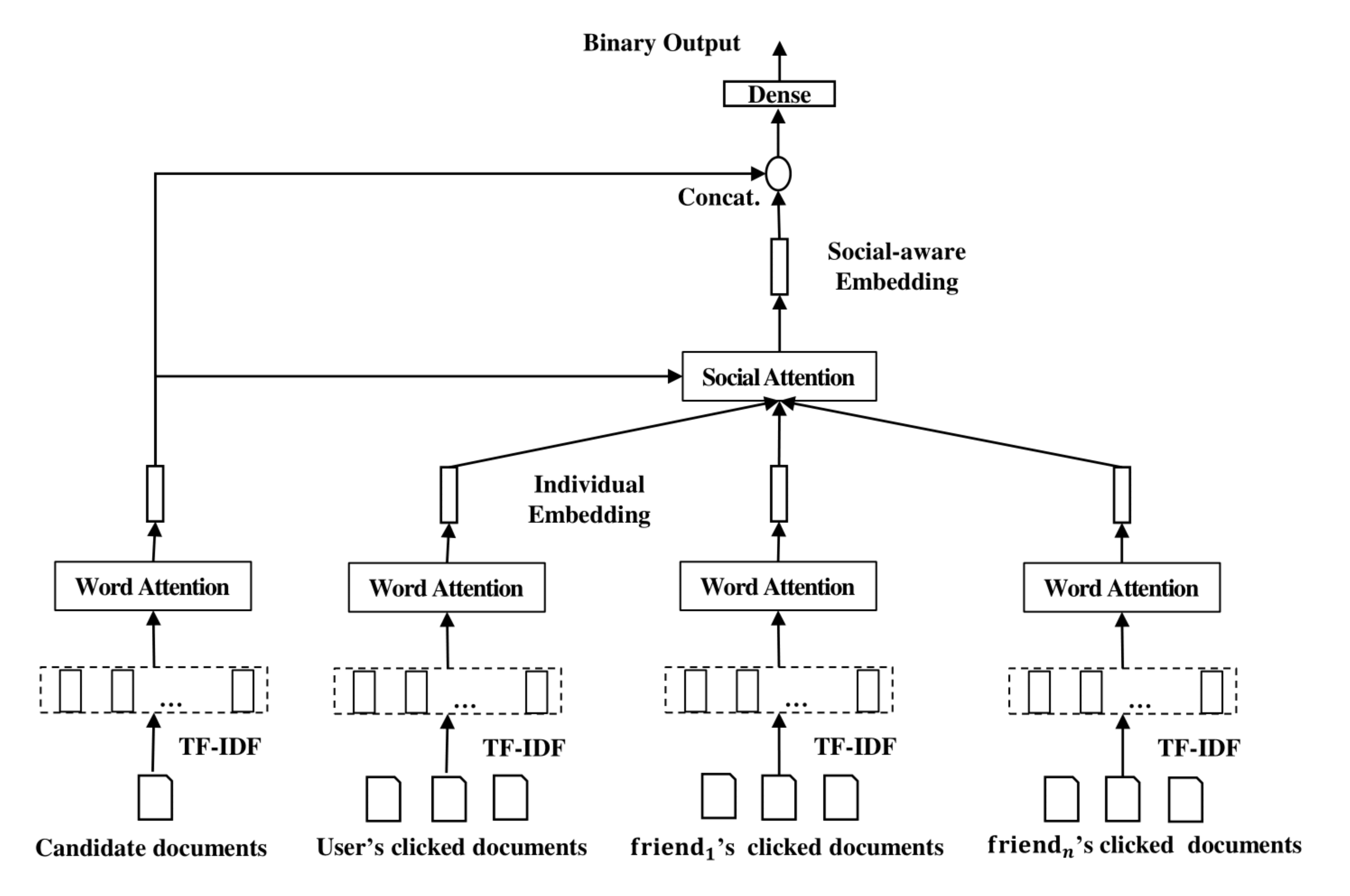}
\caption{The overview architecture of SEAN-KEYWORD. Firstly, we use TF-IDF to extract keywords to represent the candidate document and the user, respectively. Then, we apply a word attention layer for the representations of the user and the document. For the social attention layer, the model aggregates all the information from both the user and his/her friends by learning the knowledge contained in the social graph. Finally, the representation is used via the logistic regression to compute the clicking probability.}
\label{scan-model}
\end{figure}

\wenyi{In this section, we introduce the SEAN-KEYWORD model. 
The overview architecture of SEAN-KEYWORD is shown in Figure \ref{scan-model}. SEAN-KEYWORD is a content-based model for click-through rate (CTR) prediction, which takes a candidate document, i.e., the article, and a user' clicked history as inputs, and outputs the probability of the user clicking the document. It is a much faster and simpler approach for personalized representations for user and document.}

\subsection{Keywords Extraction}
\label{sec-keywords}
We first use TF-IDF to obtain several keywords from the user's reading history to capture his/her long-term interests coarsely. Since the target user is represented by all his/her reading history, it is less practical to use LSTM \citep{okura2017embedding} or CNN \citep{wang2017dynamic} to embed all the documents the users read in textual order due to the computational and space cost. Similarly, we use TF-IDF to rank different words in a document and then use the top words in the ranking list as the representation of the document. Such representations of the document and the user will be used as inputs for the later deep learning model.


\vspace{-0.13in}
\subsection{Word Attention Layer}
\label{sec-word-attention}
The word attention layer is to learn representations of a user, his/her friends, and a candidate document, respectively. We use pre-trained word embeddings as the representation of each word and fix them during training. 
Let $X = \left \{\bx_{1},...,\bx_{m} \right \}$ be word embeddings of keywords extracted from historical documents read by the target user $u$, where $m$ is the number of keywords to represent $u$.
$\bx_{k} \in \mathbb{R}^{D}$ with $D$ is the pre-defined dimension of each embedding vector. 

An attention mechanism \citep{bahdanau2014neural} is used to extract words that are important to the user's interests as well as the meaning of the document. Specifically, we define the latent representation $\bv$ of a user as:
\begin{eqnarray}
&& \bh_{k}=\tanh(\bW_w \bx_{k}+\bb_w), \\
&& \ba_{k} =  \text{Softmax}( \bu_w^\top \bh_{k}), \\
&& \bv=\sum_{k}\ba_{k} \bh_{k},
\end{eqnarray}
where the input word embedding $\bx_{k}$ is put through a non-linear layer, parameterized by a global weight matrix $\bW_w \in \mathbb{R}^{{h}\times d}$ to get a hidden representation $\bh_{k}$. $h$ is the predefined dimension of each hidden vector. $\bW_w$ and $\bb_w$ are updated day-by-day by all training samples. The context vector $\bu_w$ is randomly initialized and learned during the training process and also updated day-by-day. Moreover, the vector $\bu_w$ is applied to the user and his/her friends, which can be regarded as a virtual query, querying  informative words to the user. Then $\bv$ is computed as a weighted sum of word embeddings based on the weights, denoted as importance degrees of words interacted with the context vector $\bu_w$. We do the same process among users' friends in parallel. 

Similarly, we use another word attention module to obtain the representation of the candidate document, denoted as $\bq$.

\subsection{Social Attention Layer}
\label{sec-social}
Now we introduce how we aggregate the friends' influences to enrich the representation of a given user. 

The input to this layer is a set of users' representation vectors obtained from previous layers, including the given user's and his/her friends'. The set of users' friends' representations is denoted as $V= \left \{  \bv_{1},\bv_{2},...,\bv_{L} \right \}$. Here, $L$ is the number of the given user's selected friends, $\bv_{j} \in \mathbb{R}^h$. 
$\bv_{j}$ means the representation of friend $f_{j}$ of user $u$. 
The output is a  social-aware user's representation $\hat{\bv}$, which incorporates the information from the given user's friends. In the following steps, we present two social attention approaches, static attention, and dynamic attention, to build a social-aware representation $\hat{\bv}$ for the user $u$. 

\subsubsection{Static Attention}
The static attention based social-aware representation often performs a sum of weighted representations from friends as a general representation, while the weights are computed by certain similarity functions towards each representation of the user's friends and the user himself/herself \citep{mcpherson2001birds,ma2011recommender}.

Then the general representation from his/her friends is added to the user representation $\bv$. The equation is as follows:
\begin{eqnarray}
&& \hat{\bv} =\bv +  \sum_{j=1}^{L} \text{Softmax}(\delta (\bv, \bv_j))\bv_j,
\end{eqnarray}
where $\delta$ is the similarity functions. In this article, we give nine similarity functions, described in Table \ref{tb-sim-func}, to calculate the influence of different friends on user $u$.

\begin{table}[]
\centering

\caption{The descriptions of different similarity functions.}
\begin{tabular}{cc}
\toprule
\textbf{$\mathcal{\delta}$} & Description \\
\midrule
Cosine & $f\left ( x,y \right )=\frac{xy^\intercal}{\left \| x \right \|\left \| y \right \|}$ \\
Polynomial & $f\left ( x,y \right )=\left ( \gamma xy^\intercal + c\right )^{d}$ \\
Sigmoid & $f\left ( x,y \right )=\tanh\left ( \gamma xy^\intercal + c\right )$ \\
RBF & $f\left ( x,y \right )=\exp\left ( -\gamma \left \| x-y \right \|^{2} \right )$ \\
Euclidean & $f\left ( x,y \right )=\frac{1}{1+\left \| x-y \right \|}$ \\
Exponential & $f\left ( x,y \right )=\exp\left ( -\gamma \left \| x-y \right \|_{1} \right )$ \\
Manhattan & $f\left ( x,y \right )=\frac{1}{1+\left \| x-y \right \|_1}$ \\
GESD & $f\left ( x,y \right )=\frac{1}{1+\left \| x-y \right \|}\cdot \frac{1}{1+exp\left [ -\gamma \left ( xy^{\intercal} + c\right ) \right ]}$ \\
AESD & $f\left ( x,y \right )=\frac{1}{1+\left \| x-y \right \|} + \frac{1}{1+exp\left [ -\gamma \left ( xy^{\intercal} + c\right ) \right ]}$ \\
\bottomrule
\end{tabular}
\label{tb-sim-func}
\end{table}

\subsubsection{Dynamic Attention}
Inspired by graph attention networks~\citep{velickovic2018graph}, we present a novel method to dynamically calculate the importance of each friend to the given user.


Firstly, a shared two-layer network is applied to compute the attention $\mathcal{A}$ on the user's embedding $\bv$ with each friend's $\bv_j$ considering the document's embedding $\bq$:
\begin{equation}
e_{j}=\mathcal{A}(\bv, \bv_j, \bq)= \bu_v^\top\text{ReLU}(\bW_1\bv + \bW_2\bv_j + \bW_3\bq + \bb),
\end{equation}
where $\bW_1 \in \mathbb{R}^{h \times h}, \bW_1 \in \mathbb{R}^{h \times h}, \bW_1 \in \mathbb{R}^{h \times h}, \bb \in \mathbb{R}, \bu_v \in \mathbb{R}^h$ are model parameters, $h$ denotes the dimension of attention network and RELU is a nonlinear activation function. $e_j$ indicates the informative degree of the friend to the user. 

Secondly, we put the attention coefficients through a softmax function to get a normalized importance by
\begin{equation}
a_{j}= \frac{\exp(e_{j})}{\sum_{k \in V}\exp(e_{k})}.
\end{equation}

Thirdly, we obtain the friend's general representation $\bf$ by processing a weighted sum of these friends. Then it is added to the user's individual representation $\bv$ to get a social-aware user's representation $\hat{\bv}$:
\begin{eqnarray}
&&\bf = \sum_{j \in V}a_{j}\bv_j, \\
&& \hat{\bv} = \bv + \bf.
\end{eqnarray}

After all these operations, we obtain the representation of document $\bq$ and the social-aware representation of user $\hat{\bv}$ as:
\begin{equation}
\bo =\bq\parallel \hat{\bv}.
\end{equation}

Finally, we use a sigmoid function to predict the probability that the target user $u$ will read the candidate document $d$ over the model. For SEAN-KEYWORD, we also adopt the binary cross-entropy loss to train the whole model, the same as SEAN-END2END.

%


\section{Complexity Analysis and Discussion}
In this section, we give the complexity analysis on two variants of SEAN and discussing the differences and similarities of the two models.

\subsection{Complexity Analysis}
\label{sec-complexity}
\subsubsection{For SEAN-END2END}
For the word level, the time complexity is linear to the number of tokens in the training data set, which is $M \cdot I \cdot J$, where $M$ is the number of training samples,  $I$ is the maximum number of sentences, and $J$ is the maximum number of tokens in a sentence.
It is also linear to the number of kernels $K$, the numbers of hidden vectors $h$, the filter size $r$, and the convolutional window size $g$. Since we use fixed-sized word embeddings, the large number of words do not contribute to our time cost.
For the sentence level, the time cost of the GRU layer is linear to the maximum number of sentences $I$ and the number of parameters in the GRU cell $P_{GRU}$.
For both word-level and sentence-level attentions, the cost is linear to the square number of hidden dimension $h^2$, the number of selected friends $L$, and the times of trials of attention $B$. Note that the number of selected friends $L$ and the times of trials of attention $B$ are the same as the search depth $L$ and beam width $B$  introduced in Section~\ref{sec-social-exploration}.
Moreover, for the fully connected layer, the parameter is linear to $h$.
Therefore, the overall time complexity is 
$O(M \cdot I \cdot J \cdot (K\cdot g \cdot r \cdot D  + h \cdot r \cdot K + h^2\cdot L \cdot B) + M \cdot I \cdot (P_{GRU} + h^2\cdot L \cdot B) )$.

\subsubsection{For SEAN-KEYWORD}
For the word attention layer to represent the target user, the time cost is linear to the number of tokens in the training data set, which is $O(M \cdot m \cdot L \cdot D \cdot h)$, where $M$ is the number of training samples,  $m$ is the number of keywords for the target user, $L$ is the number of friends for the target user, $D$ is the dimension of word embeddings, $h$ is the dimension of the hidden vector.
For the social attention layer, the time cost is linear to the number of attended friends $L$ and the square number of hidden dimension $h^2$, which is $O(M \cdot L \cdot h^2)$. 
To represent the candidate document, the time complexity is linear to the number of training samples as well as the keywords extracted for the candidate document, which is $O(M \cdot n \cdot D \cdot h)$, where $n$ is the number of keywords for candidate document. Therefore, the overall complexity is 
$O(M \cdot m \cdot L \cdot D \cdot h + M \cdot L \cdot h^2 + M \cdot n \cdot D \cdot h)$.

\subsection{Discussion on the two variants of SEAN}
\wenyi{SEAN-END2END and SEAN-KEYWORD are named according to the techniques of using attentions among users and documents. In this section, we provide a unified view of the two versions of SEAN and show that their major difference lies in whether we use a deep learning based method to extract user's interested points in the documents. 
Recall that SEAN-END2END is an end-to-end model that combines the three processes, pointing users' interested points, constructing personalized document representation, and predicting users' clicking. 
However, SEAN-KEYWORD uses non-deep learning based methods to first find keywords coarsely, a much faster and scalable approach to attending users' interests, then applies the deep learning based approaches for representation and prediction. In general, SEAN-END2END achieves higher effectiveness while SEAN-KEYWORD achieves higher efficiency.}

\section{Experiments}
\label{sec-exp}
In this section, we present our experimental results. We firstly introduce dataset description, evaluation metrics, baseline comparison, and experimental settings. Then we show the performance comparisons between our models and baselines, followed by extensive study of our models, including different social exploration methods, ablation study, hyper-parameters, and scalability analysis.

\subsection{Dataset Description}
\label{sec-dataset}
We build two datasets, Steemit-English and Steemit-Spanish from the decentralized social platform, Steemit.
Steemit is a blogging and social networking platform that uses the Steem blockchain to reward creators and consumers. 
Most of the modern content distribution platforms are already using recommendation systems to recommend contents to users, which can be biased if we collected data from them for our evaluation. 
Different from them, the contents and user clicks are not manipulated by the Steemit platform.
We retrieve the commenting activities of users (consumers) from June \nth{2}, 2017 to July \nth{6}, 2018. 
Two datasets are constructed based on social communities using English and Spanish respectively.

We form a sample as a triplet with three elements: a given user, a document, and a label $1/0$. We form the positive samples by the documents in which users have made comments. We treat messages that users' friends have made responses but the users themselves do not as negative samples.  Since we collect users' activity information from their comment logs, it is natural that the number of users who made comments on this platform is not too much. 
The statistics of the two datasets are shown in Table \ref{table-data-descript}. 

\begin{table}[]
    \centering
    \caption{Statistics of the two datasets.}
    \label{table-data-descript}
\begin{tabular}{ccc}
\toprule
 & \textbf{Steemit-English} & \textbf{Steemit-Spanish} \\
\midrule
Duration (days)  & 370  & 126 \\
\# Consumers & 7,242 & 1,396 \\
\# Creators & 44,220 & 4,073 \\
\# Relations & 273,942 & 25,593 \\
\# Documents & 177,134 & 14,843 \\
Avg. word per document & 290 & 509\\
\# Logs & 220,909 & 20,893\\
\# Samples & 684,752 & 92,236\\
\bottomrule
\end{tabular}
\end{table}

\subsection{Evaluation Metrics}
\label{sec-metrics}
To evaluate the recommendation quality of the proposed approach, we use the following metrics: Area under the Curve of ROC (AUC) and F1 for consumer satisfaction and the Gini coefficient for creator equality, where Gini coefficient is defined as:
\begin{equation}
    \text{Gini} = \frac{\sum_{i=1}^{n}(2i-n-1)x_i}{n\sum_{i=1}^{n}x_i},
\end{equation}
where, $x$ is an observed value, $n$ is the number of values observed, and $i$ is the rank of values in ascending order.
To measure the performance of models considering both creators and consumers, we calculate the harmonic mean of F1 and (1-Gini), denoted as C\&C:
\begin{equation}
    \text{C\&C} = \frac{2\times (1-\text{Gini})\times \text{F1}}{(1-\text{Gini})+\text{F1}}.
\end{equation}
Since we train and test day-by-day, we compare a model's quality by the average of each metric during the whole period. For AUC, F1, and C\&C, the larger, the better. For Gini, the smaller, the better.

\subsection{Baselines}
\label{sec-baselines}
We compare our model with following baselines.

\textbf{LR} \citep{van2013deep} is the simplest word-based model for CTR prediction. We use TF-IDF to extract keywords for a user's clicked historical documents and the new incoming document and feed them to a logistic regression model to predict the label. 

\textbf{LibFM} \citep{rendle2012factorization} is a state-of-the-art feature-based factorization model and widely used in CTR prediction. In this article, we use the same features as LR and feed them to LibFM. LibFM treats a user's features and a document's features separately for the factorization.


\textbf{DKN}~\citep{wang2018dkn} learns representations of documents and users. In DKN, it obtains a set of embedding vectors for a user's clicked historical documents. Then an attention is applied to automatically match the candidate document to each piece of his/her clicked documents, and aggregate them with different weights. 
Here, we only use DKN's base model without the knowledge graph information.

\textbf{NCF} \citep{he2017neural} is short for Neural network based Collaborative Filtering. It is a deep model for recommender systems that uses a multi-layer perceptron (MLP) to learn the user$-$item interaction function. It ignores the content of news and uses the comment counting information as input.

\textbf{SAMN}~\citep{chen2019social}, Social Attentional Memory Network, is a collaborative filtering model that employs the attention-based memory module to attend to a user's one-hop friends' vectors. The attention adaptively measures the social influence strength among friends. 

\textbf{SEAN-END2END \& SEAN-KEYWORD}, are two versions of SEAN proposed in the article. If without any clarification, we use F1 score as the exploitation value.

\subsection{Experimental Settings}
\label{sec-exp-settings}
For our framework, we use pre-trained word embeddings for the document and fix them during training for both two versions of SEAN.
\wenyi{In SEAN-END2END, for the word-level representation in the CNN layer, the filter number is set as 50 for each of the window sizes ranging from 1 to 6.
The hidden vector size is set to 128 for both GRU layers and dense layers. 
The beam width $B$ is set to 3 and $\lambda$ is set to 1. 
The search depth $L$ is set to 10.
We train the model for 6 epochs every day. 

In SEAN-KEYWORD, We set the number of keywords retrieved by TF-IDF to 90 to represent the document, and 200 to depict the user’s reading history on both Steemit-English and Steemit-Spanish. The size of the hidden vector is set to 64 for all dense layers. The epoch is set to 5 for everyday training. 

The key parameter settings for baselines are as follows. 
For the keyword extraction in LR and LibFM, we set the number of keywords for document and user's historical readings as 20 and 90. 
For DKN, the length of the document embedding is set to 200. Due to the limitation of memory, we use a user's latest 10 clicked documents to represent the user.
For the CF based methods, NCF and SAMN, the embedding size of users and the documents are all set to 128.

The above settings are for fair consideration. Each experiment is repeated five times, and we report the average and standard deviation as results.
The data every day is split to 9:1 for training and validation. We train the model from the data of past $t$ days and test it by using the data on $t+1$-th day. The whole model is implemented on Keras 2.0 with Tensorflow 1.12 as the backend, based on CUDA 7.5 using a single GPU, GeForce GTX 1080 with 8GB VRAM.
}


\subsection{Comparison with Baselines}
\label{sec-vs-baselines}
Table \ref{tb-performance} reports the results on Steemit-English and Steemit-Spanish datasets.  For consumers, SEAN-END2END improves F1 by above 5 percentage points and AUC by near 3 percentage points compared with the best content-based model DKN on Steemit-English and improves F1 by 1.7 percentage point and AUC by near 3 percentage point on Steemit-Spanish. 
This proves that our model can best consider consumer's interests and recommend the most interesting contents to them. 
LR and LibFM perform much worse because these two models ignore the word order information and consequently generate worse document and user representations. 
Moreover, compared with CF-based models (NCF and SAMN), SEAN-END2END can also outperform them significantly. This result shows that CF methods cannot work well in this recommendation scenario since the documents on Steemit is highly time-sensitive, and the content should be considered for the recommendation.
Besides, from the comparison with the SAMN, we can see that our strategy to incorporate social information is more effective than SAMN. 

For creators, the Gini coefficients of the content-based models are smaller than those of the CF-based models. The result proves our aforementioned claim that CF methods are more likely to suffer from Matthew's effect since CF-based models intend to use global behavioral information to promote popular documents on the social platform. 
The Gini coefficients of SEAN-END2END and DKN are comparable, which shows that under the premise of the quality of recommendation for consumers, our algorithm can also encourage creator's equality which may further encourage creators to stay on the platform to keep publishing their innovative contents.

From the harmonic mean C\&C results, we observe that SEAN-END2END performs best on both datasets. 
This demonstrates that the social exploration mechanism can have a good balance on optimizing between consumer satisfaction and creator equality.

\wenyi{For SEAN-KEYWORD, it is worse than DKN and SEAN-END2END for both consumers (AUC, F1 score) and creators (Gini coefficients). However, it outperforms NCF, SAMN, LR, and LibFM. The reason would be that both DKN and SEAN-END2END use more complicated components (LSTM, CNN) to encode the document. Comparing two SEAN models, it indicates that an end-to-end model to get user's interests for the final prediction can get better performance than first retrieving user's keywords then prediction.}

\begin{table}[]
    \centering
    \caption{
    Comparison of different methods on Steemit. The best results are highlighted in boldface.
    }
    \label{tb-performance}
\resizebox{0.95\textwidth}{!}{
\begin{tabular}{cccccc}
\toprule
Dataset & Model & AUC & F1 & Gini & C\&C \\
\midrule
\multirow{7}{*}{English} & NCF & 52.83$\pm$0.13 & 42.14$\pm$0.21 & 66.04$\pm$0.25 & 37.71$\pm$0.22 \\
 & SAMN & 53.05$\pm$0.35 & 42.28$\pm$0.45 & 65.98$\pm$0.21 & 37.80$\pm$0.28 \\
 & LR & 52.89$\pm$0.07 & 34.50$\pm$0.11 & 62.89$\pm$0.11 & 35.86$\pm$0.11 \\
 & LibFM & 50.01$\pm$0.12 & 40.43$\pm$0.22 & 66.42$\pm$0.13 & 36.79$\pm$0.16 \\
 & DKN & 62.71$\pm$0.22 & 42.85$\pm$0.45 & 62.29$\pm$0.26 & 40.22$\pm$0.33 \\
 & SEAN-KEYWORD & 55.59$\pm$0.39 & 42.96$\pm$0.45 & 64.00$\pm$0.25 & 39.17$\pm$0.32 \\ 
 & SEAN-END2END & \textbf{65.57$\pm$0.17} & \textbf{47.69$\pm$0.46} & \textbf{61.78$\pm$0.24} & \textbf{42.43$\pm$0.33} \\
 \hline
\multirow{7}{*}{Spanish} & NCF & 50.46$\pm$0.21 & 35.02$\pm$0.26 & 58.13$\pm$0.34 & 38.14$\pm$0.29 \\
 & SAMN & 51.10$\pm$0.24 & 35.24$\pm$0.31 & 58.29$\pm$0.32 & 38.20$\pm$0.31 \\
 & LR & 53.15$\pm$0.06 & 36.50$\pm$0.29 & 55.84$\pm$0.09 & 39.97$\pm$0.14 \\
 & LibFM & 47.71$\pm$0.30 & 22.37$\pm$0.33 & 56.50$\pm$0.21 & 29.55$\pm$0.26 \\
 & DKN & 57.02$\pm$0.39 & 41.27$\pm$0.45 & \textbf{53.98$\pm$0.25} & 43.52$\pm$0.32 \\
 & SEAN-KEYWORD & 55.83$\pm$0.29 & 41.04$\pm$0.34 & 58.19$\pm$0.23 & 41.42$\pm$0.32 \\
 & SEAN-END2END & \textbf{59.98$\pm$0.34} & \textbf{42.99$\pm$0.37} & 53.99$\pm$0.23 & \textbf{44.46$\pm$0.28}\\
 \bottomrule
\end{tabular}
}
\end{table}

\subsection{Different Strategies in Social Exploration }
\label{sec-vs-exploration}

In the experiment, we evaluate the performance of each exploitation-exploration method using the Steemit-English dataset.
\wenyi{Since the social exploration strategy is used in SEAN-END2END and SEAN-KEYWORD without any difference, we only show the results with SEAN-END2END for simplicity.}
``Random Select'' is the model that randomly selects a set of users on the social platform as the target user's friends. 
``Random Walk'' is the model that uses a stochastic process, moving from a node to another adjacent node. These two models are both using a random based strategy to explore.
As shown in Table \ref{tb-exploration-methods}, MCTS based models have better F1 than random based models, because the exploitation mechanism can help the model find more relevant friends. SEAN-RS-F1 performs the best on F1 because this model tends to explore friends of higher quality continuously by directly using the recommendation feedbacks. The F1 performance of SEAN-SPR and SEAN-DPR are compatible, while both are worse than the others since SEAN-SPR uses the static social network and SEAN-DPR only uses the daily comment network formed by consumer-creator connections, both missing some information. 
Moreover, MCTS based models also outperform random based models on C\&C, which indicates that our model can improve the recommendation quality for consumers even though slightly hurts the equality. Specifically, SEAN-Payout has the highest C\&C which indicates that using payout, the rewards given by Steemit as the exploitation value, is more suitable to select friends on this platform. Meanwhile, it further verifies the decentralized nature of this platform.     

\wenyi{$\epsilon$-Greedy is SEAN with the strategy $\epsilon$-Greedy mentioned in Section \ref{sec-greedy-exploration}. Compared with MCTS based models, it performs worse on F1 while better on Gini. It is because that $\epsilon$-Greedy has a certain probability to randomly select from neighbors, a more random way to explore new friends. The C\&C shows that MCTS is a smarter and more dynamic way to balance exploitation and exploration compared with $\epsilon$-Greedy.}
\begin{table}[]
    \centering
    \caption{Comparison of social exploration methods.}
    \label{tb-exploration-methods}
{
\begin{tabular}{cccc}
\toprule
Models & F1 & Gini & C\&C \\ 
\midrule
Random Select & 42.48$\pm$0.38 & \textbf{59.13$\pm$0.22} & 41.09$\pm$0.28 \\
Random Walk & 45.05$\pm$0.39 & 60.98$\pm$0.09 & 41.77$\pm$0.20 \\ \hline
SEAN-RS-F1 & \textbf{47.69$\pm$0.46} & 61.78$\pm$0.24 & 42.43$\pm$0.33 \\
SEAN-SPR & 45.99$\pm$0.35 & 60.90$\pm$0.32 & 42.27$\pm$0.33 \\
SEAN-DPR & 45.96$\pm$0.44 & 60.98$\pm$0.22 & 42.21$\pm$0.29 \\
SEAN-Payout & 46.26$\pm$0.36 & 60.65$\pm$0.40 & \textbf{42.53$\pm$0.37} \\ \hline
\wenyi{$\epsilon$-Greedy} & 43.95$\pm$0.42 & 59.63$\pm$0.19 & 42.08$\pm$0.20 \\ 
\bottomrule
\end{tabular}
}
\end{table}

\subsection{Compare Running Time of Two SEAN Models}
\label{sec-two-sean-comparison}
\begin{figure}[]
\caption{Time Comparison between SEAN-END2END and SEAN-KEYWORD.}
\label{fig-time-compare}
\centering
\includegraphics[width=0.65\textwidth]{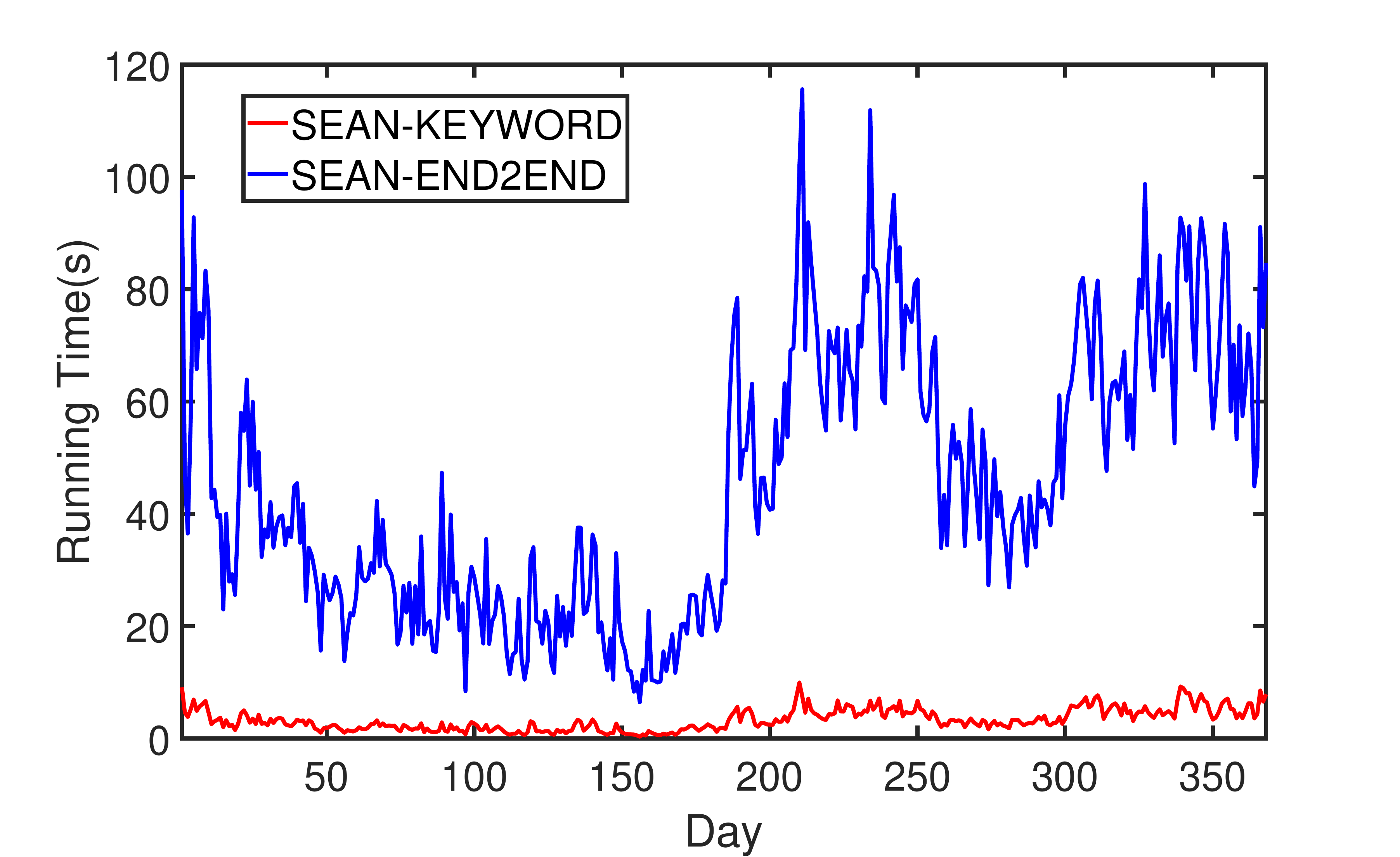}
\end{figure}

    \wenyi{We give the running time comparison of two SEAN models. As shown in Figure \ref{fig-time-compare}, the running time of SEAN-KEYWORD is much faster than SEAN-END2END since there are no complex modules in SEAN-KEYWORD, e.g., CNN or LSTM, expect for simple attention modules (word attention to social attention). Another reason should be that using top-K keywords is efficient than an end-to-end manner to point out users' reading keywords. 
    In other words, SEAN-KEYWORD is more suitable to handle very large-scale data in real-world industrial scenarios.}

\subsection{Model Ablation Study}
\label{sec-ablation}

\subsubsection{Variants of Socialization}
\label{sec-as-socializaton}
We compare variants of the socialization component in SEAN in terms of the following aspects to demonstrate the efficacy of the framework design: the use of social connections, the use of social attention, the use of friend exploration. Same as in Section~\ref{sec-vs-exploration}, we only conduct experiments with SEAN-END2END for simplicity, and the results are shown in Table \ref{tb-social-ablation}.

For the consumer side, we can conclude as follows.
\begin{itemize}
    \item Without any social information means that we are using a pure personalization model for each user. This will decrease F1 by 5 percentage points. This confirms the efficacy of using social information in the SEAN-END2END.
    \item We also replace social attention with simple averaging friends' representation vectors. This results in a loss of F1 by near 3 percentage points. In other words, it demonstrates the effectiveness of weighing different social influences from friends on recommendation performance.
    \item We use each user's first-order (one-hop) friends for socialization. This is also worse than SEAN-END2END with exploring high-order friends, which proves the importance of exploring friends for recommendation.
\end{itemize}

For the creator side, the Gini coefficient of SEAN-END2END w/o social is the lowest one, followed by SEAN-END2END with one-hop friends. The reason is that without using any social information, users are not influenced by other users' reading histories, thus cutting off the spread of popular documents. Besides, using the first-order connections is worse than using high-order social information.

\begin{table}[]
    \centering
    \caption{Comparison of different variants for socialization.}
    \label{tb-social-ablation}
{
\begin{tabular}{cccc}
\toprule
Variants & F1 & Gini & C\&C \\
\midrule
w/o social & 42.40$\pm$0.30 & \textbf{58.56$\pm$0.43} & 41.91$\pm$0.35 \\
w/o social attention & 44.79$\pm$0.17 & 62.22$\pm$0.36 & 41.98$\pm$0.23 \\
one-hop friends & 43.08$\pm$0.16 & 60.85$\pm$0.25 & 41.04$\pm$0.20 \\
\hline
SEAN-END2END & \textbf{47.69$\pm$0.46} & 61.78$\pm$0.24 & \textbf{42.43$\pm$0.33}\\
\bottomrule
\end{tabular}
}
\end{table}

\subsubsection{Variants of SEAN-END2END}
\label{sec-as-end2end}
We further compare different components in the hierarchical document representation of our SEAN-END2END model to demonstrate the efficacy of the RS model design. The results are shown in Table \ref{tb-ablation}. Specifically, we test how CNN for word-level and GRU for sentence-level encoding affect the performance. The usage of CNN and GRU brings about 2 percentage points to gain on F1 respectively. Without using GRU and CNN decreases F1 by more than 3 percentage points. For the creator side, for models without GRU and/or CNN components, Gini drops within 2 percentage points while F1 also drops. The best result of C\&C indicates that our model can obviously improve consumers' satisfaction without hurting equality too much.
\begin{table}[]
    \centering
    \caption{Comparison of different variants of SEAN-END2END.}
    \label{tb-ablation}
{
\begin{tabular}{cccc}
\toprule
Variants & F1 & Gini & C\&C \\
\midrule
w/o CNN  & 45.25$\pm$0.22 & \textbf{59.98$\pm$0.26} & 42.58$\pm$0.21 \\
w/o GRU  & 45.07$\pm$0.31 & 60.47$\pm$0.14 & 42.12$\pm$0.27 \\
w/o CNN \& GRU  & 44.08$\pm$0.26 & 60.06$\pm$0.20 & 41.91$\pm$0.25 \\
\hline
SEAN-END2END & \textbf{47.69$\pm$0.46} & 61.78$\pm$0.24 & \textbf{42.43$\pm$0.33}\\
\bottomrule
\end{tabular}
}
\end{table}

\subsubsection{Variants of SEAN-KEYWORD}
\label{sec-sean-variant}
\wenyi{We conduct experiments on a variety of social attention approaches on Steemit-English. 
As results shown in Table \ref{tb-similarity}, the dynamic attention can get the best F1 and the RBF similarity can get the best Gini. Besides, RBF achieves the best performance on C\&C.
In General, the results of dynamic attention, cosine similarity, polynomial similarity, sigmoid similarity are comparable and better than others, which indicate they are more suitable to compute the influence of different friends.}

\begin{table}[]
\centering
\caption{Comparison of various similarity functions in SEAN-KEYWORD.}
\label{tb-similarity}
\resizebox{0.95\textwidth}{!}{
\begin{tabular}{cccccc}
\toprule
Attention Type & Similarity Function & Parameter & F1 & Gini & C\&C \\
\midrule
Dynamic & Self-Attention & - & \textbf{42.96} & 64.00 & 39.17 \\
\hline
\multirow{9}{*}{Static} & Cosine & - & 42.35 & 63.70 & 39.09 \\
 & Polynomial & $\gamma = 0.5, d=2, c=1,$ & 42.61 & 63.66 & 39.23 \\
 & Sigmoid & $\gamma = 0.5, c = 1$ & 42.80 & 63.68 & \textbf{39.29} \\
 & RBF & $\gamma = 0.5$ & 34.10 & \textbf{61.75} & 36.06 \\
 & Euclidean & - & 34.24 & 62.00 & 36.02 \\
 & Exponential & $\gamma=0.5$ & 34.21 & 61.85 & 36.07 \\
 & Manhattan & - & 35.20 & 62.11 & 36.50 \\
 & GESD & $\gamma =0.5, c = 0.1$ & 35.02 & 61.97 & 36.46 \\
 & AESD & $\gamma =0.5, c = 0.1$ & 33.71 & 61.86 & 35.79 \\
 \bottomrule
\end{tabular}
}
\end{table}

\subsection{Hyper-parameter Sensitivity}
\label{sec-param-sens}
SEAN involves many hyper-parameters, including SEAN-END2END and SEAN-KEYWORD. In the following experiments, except for the parameter being tested, all other parameters are set as introduced in Section \ref{sec-exp-settings} if we do not point out. The parameter sensitivity is done by using the samples from Steemit-English during the first 100 days. 


\subsubsection{For Friend Selection}
We test the hyper-parameters in MCTS. Since the friend selection parts in SEAN-END2END and SEAN-KEYWORD are the same, we test only on SEAN-END2END for simplicity. 

\begin{figure}[]
    \caption{Hyper-parameter sensitivity analysis on Steemit-English for Socialization.}
    \label{fig-param}
    \subfigure[Search depth $L$.]{\includegraphics[width=1\textwidth]{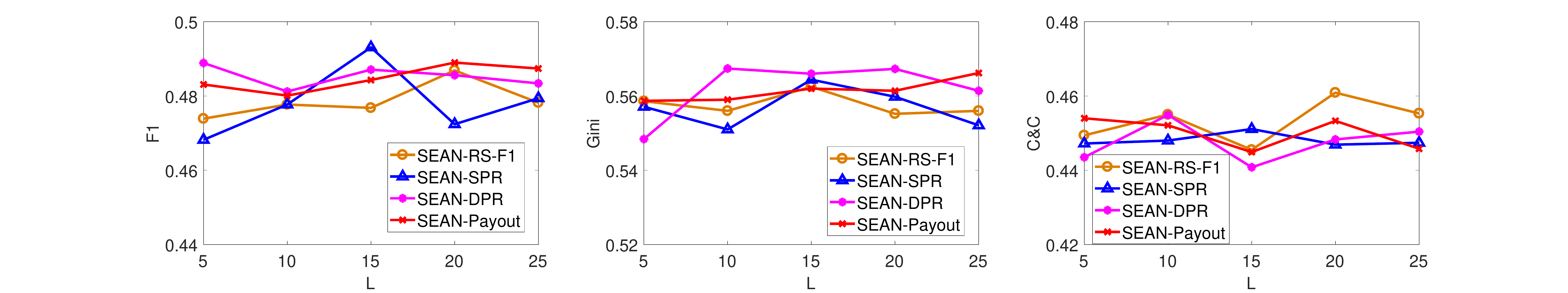}\label{fig-ps-a}}
\\
    \subfigure[Beam width $B$.]{\includegraphics[width=1\textwidth]{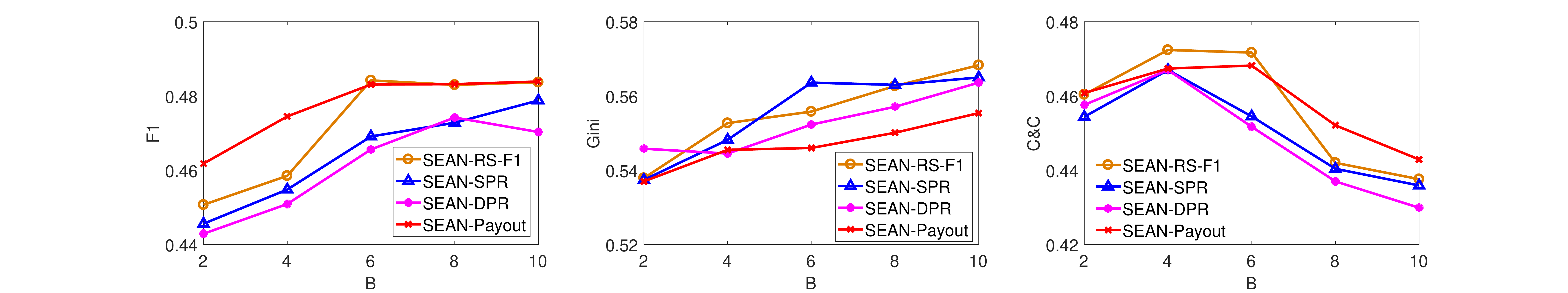}\label{fig-ps-c}}
 \\
    \subfigure[Trade-off $\lambda$.]{\includegraphics[width=1\textwidth]{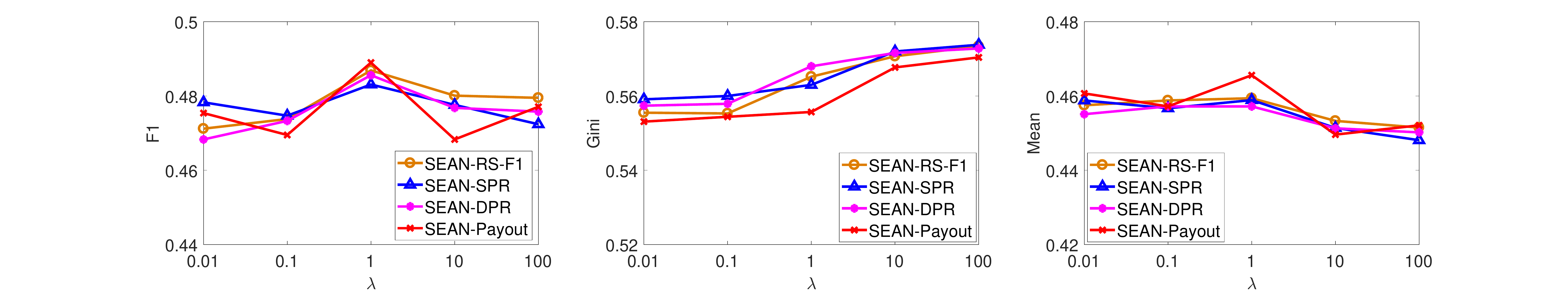}\label{fig-ps-b}}
\end{figure}

\begin{figure}[]
    \caption{Hyper-parameter sensitivity analysis on Steemit-English for $\epsilon$-Greedy.}
    \label{fig-ps-i}
    \includegraphics[width=1\textwidth]{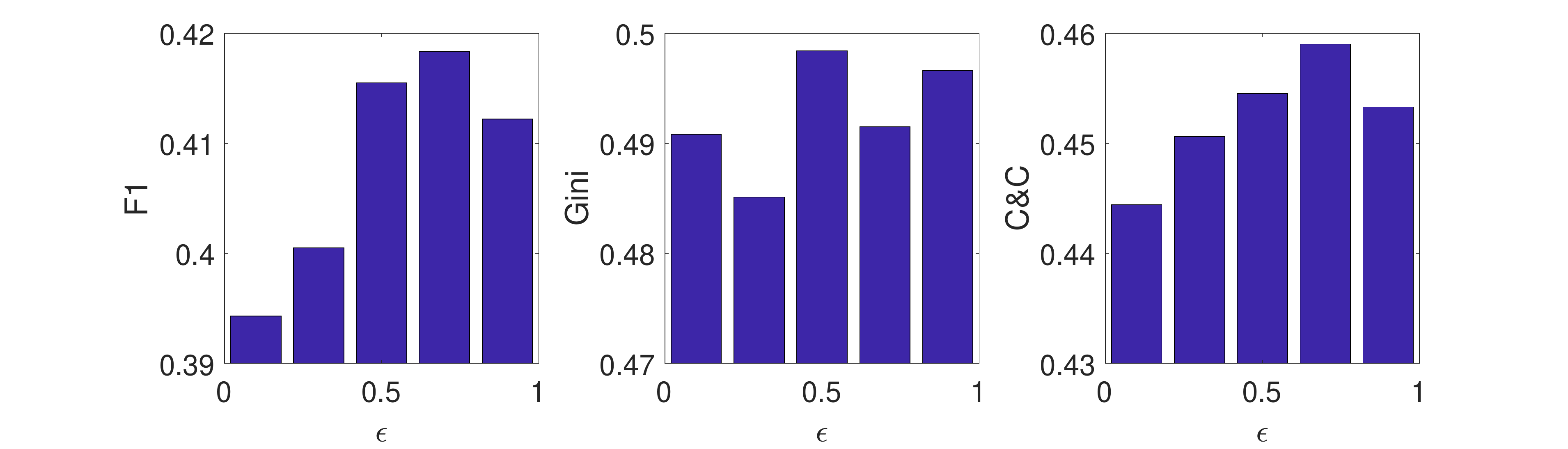}
\end{figure}

\begin{itemize}
    \item \textbf{ Search Depth $L$.}
We test the influence of search depth $L$ for four proposed models with $L= 5, 10, 15, 20, 25$. The results are shown in Figure \ref{fig-ps-a}. Given the best settings shown in Section \ref{sec-exp-settings}, changing $L$ from 5 to 25 does not affect both F1 and Gini a lot compared to the beam width $B$. This may indicate that given the Steemit network and the prediction F1 score, using a small number of friends can already cover most of the friends to explore while increasing $B$ will force the exploration to find more neighbors.
    \item \textbf{ Beam width $B$.}
We investigate the influence of the beam width $B$ (number of paths) by setting $B$ ranging from 2 to 10. The results are shown in Figure \ref{fig-ps-c}. 
We can see that F1 increases as the beam width grows since there are more selected friends that are helpful for the user. While with a continuing increase of $B$, F1 tends to be flat since the overlapping of friends selected from each path also increases. Meanwhile, Gini continuously increases when the beam width increases, and thus C\&C appears to be best only when $B$ is 4. 
    \item \textbf{ Trade-off constant  $\lambda$.}
The choice of the trade-off constant $\lambda$ is set to be $\lambda\in\left \{0.01, 0.1, 1, 10, 100\right \}$. We can see in Figure \ref{fig-ps-b}, the best F1 is at $\lambda = 1$ for all approaches. It indicates that both exploration and exploitation are important to better select friends. Besides, Gini is less influenced by $\lambda$. 
    \item \textbf{$\epsilon$-Greedy.}
    \wenyi{We also test the influence of different $\epsilon$ for using $\epsilon$-Greedy algorithm in SEAN-END2END to explore new friends. Here, we set $\epsilon$ with $\left \{0.1, 0.3, 0.5, 0.7, 0.9 \right \}$. The results are shown in Figure \ref{fig-ps-i}. The best C\&C is obtained when $\epsilon=0.7$. When $\epsilon = 0.1$, the F1 score is much worse since it has less probability to explore new friends, further demonstrating the importance of balancing exploitation and exploration.}
\end{itemize}

\subsubsection{For SEAN-END2END}
\begin{figure*}[]
 \caption{Hyper-parameter sensitivity analysis on Steemit-English for SEAN-END2END.}
    \label{app-fig-param}
    \subfigure[Hidden Size $h$.]{\includegraphics[width=1\textwidth]{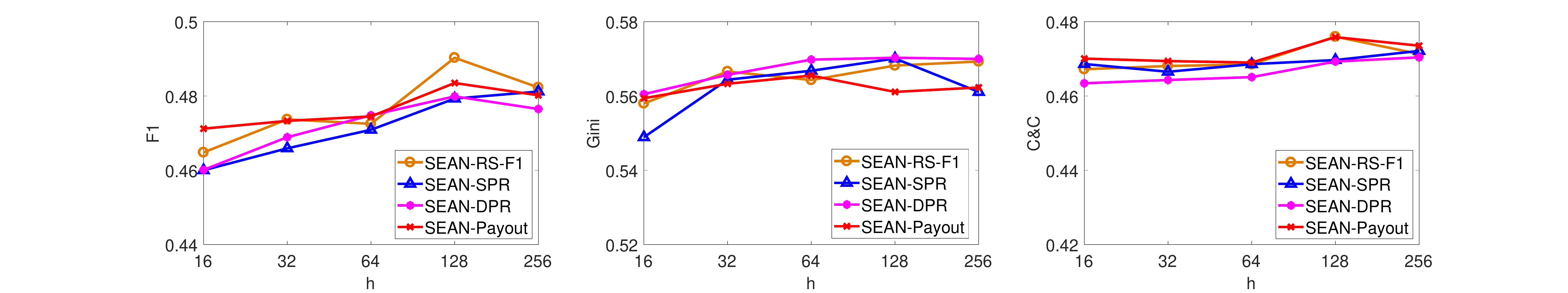}\label{fig-ps-d}}\\
    \subfigure[Number of kernels $K$ in CNN.]{\includegraphics[width=1\textwidth]{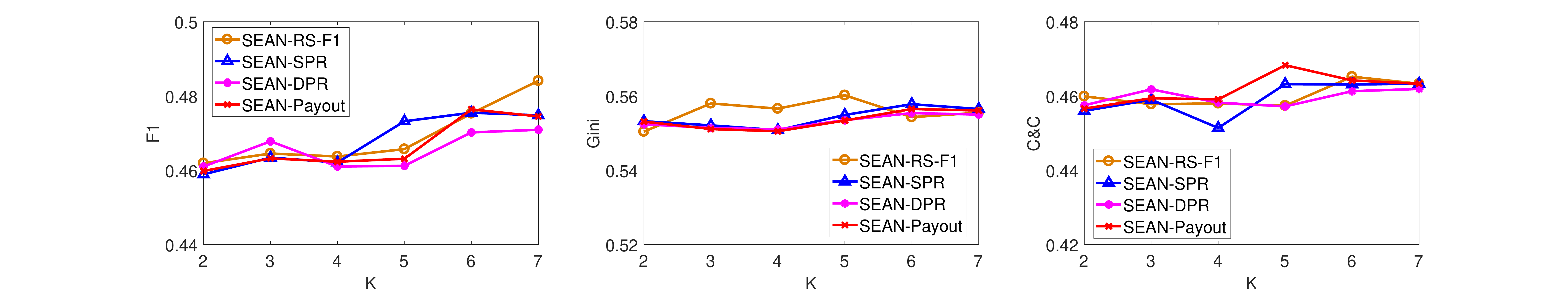}\label{fig-ps-e}} \\
\vspace{-0.1in}
    \subfigure[Filter size $r$.]{\includegraphics[width=1\textwidth]{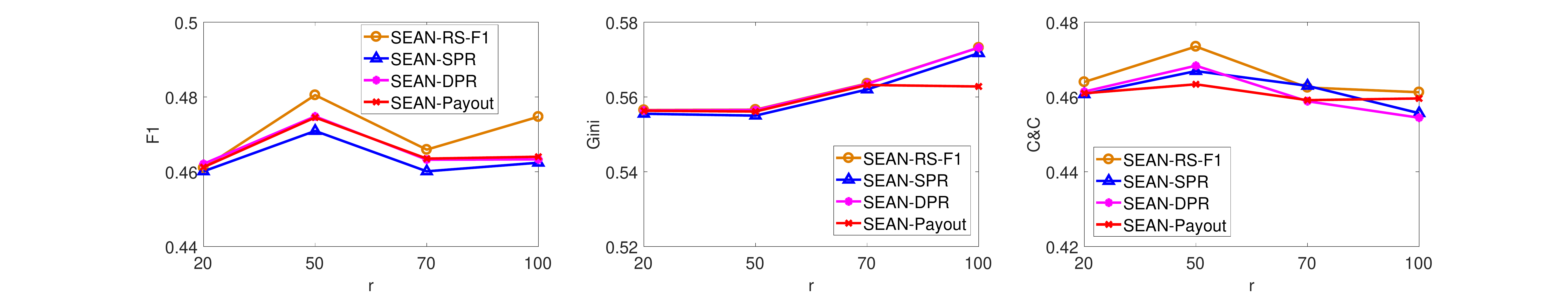}\label{fig-ps-f}}
\vspace{-0.1in}
   
\end{figure*}

\begin{itemize}
    \item \textbf{Hidden Size and User Embedding Size $h$.}
    We first investigate how the hidden size $h$ affect the performance by testing $h$ in set $\left \{ 20,50,70,100 \right \}$. The results are shown in Figure \ref{fig-ps-d}, from which we can observe that all four models obtain best F1 when $h=128$. Changing $h$ does not affect too much Gini scores. The trend of C\&C is similar to the trend of F1, also getting the best result when $h=128$.
    \item \textbf{The number of kernels $K$ and sizes of filters $r$.}
    We investigate the number of kernels $K$ and the choice of filter sizes $r$ for CNN in SEAN. As shown in Figure \ref{fig-ps-e}, the F1 score generally increases as the number of kernels $K$ with different convolutional windows $g$ gets larger, since more kernels can capture long-distance patterns in sentences. Due to the limitation of time and memory, we do not further enlarge the $K$.  Meanwhile, the influence of $K$ on Gini is smaller than on F1. SEAN-F1-RS performs best on F1 while performs worst on Gini.  We can get the best C\&C for all the proposed models except SEAN-RS-F1 when $K=6$.
    Likewise, we can observe similar rules for the filter size $r$, shown in Figure \ref{fig-ps-f}: a small filter size cannot capture more local patterns in sentences, while a large filter size may easily suffer from overfitting. The Gini increases with the increasing of filter size $r$. The best C\&C results are obtained when $r=50$.
\end{itemize}

\subsubsection{For SEAN-KEYWORD}
\begin{figure}[]
\centering
    \subfigure[\# of user keywords $m$.]{\includegraphics[width=0.85\textwidth]{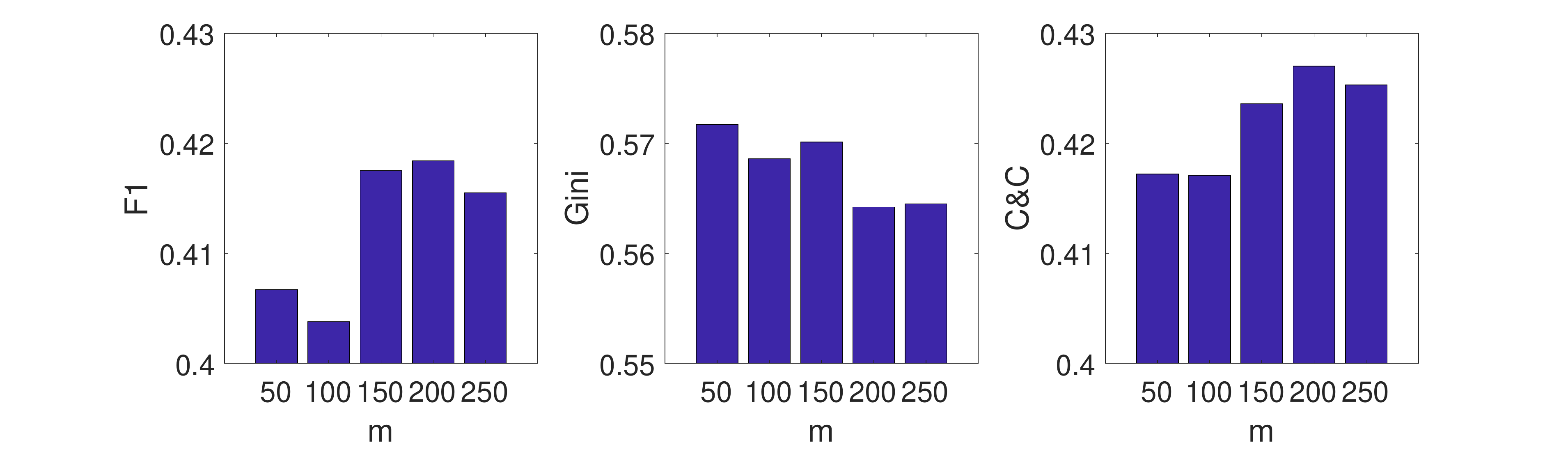}\label{fig-ps-g}}
\\
    \subfigure[\# of document keywords $n$.]{\includegraphics[width=0.85\textwidth]{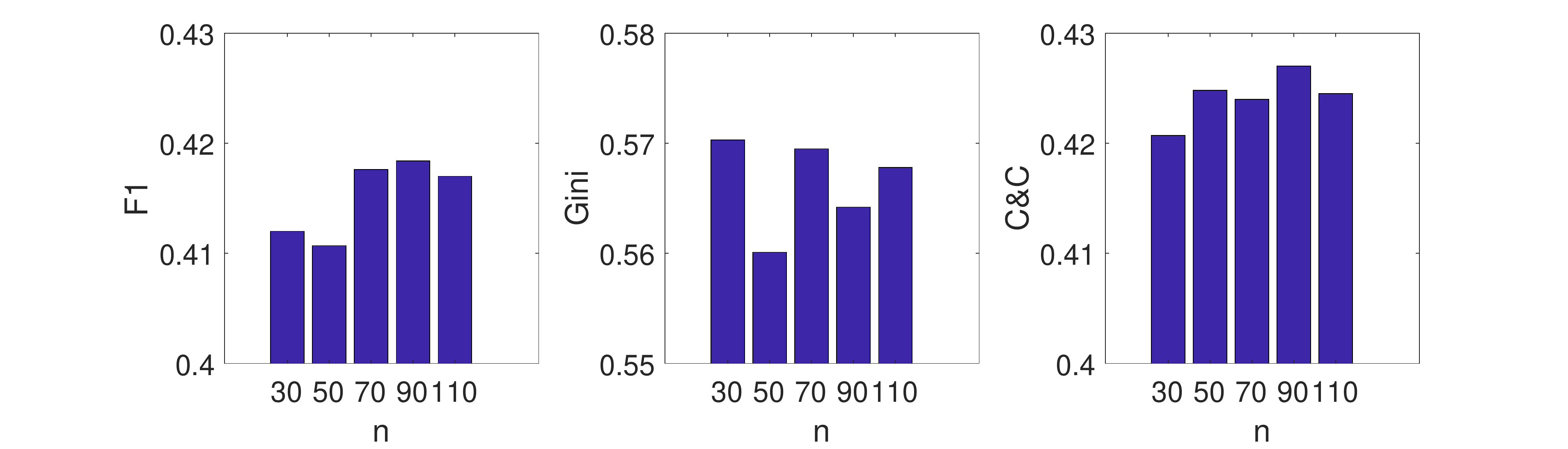}\label{fig-ps-h}}
    \caption{Hyper-parameter sensitivity analysis on Steemit-English for SEAN-KEYWORD.}
\end{figure}

\begin{itemize}
    \item \textbf{\# of user keywords $m$}. We test $m$ in $\left \{50, 100, 150, 200, 250 \right \}$ for SEAN-KEYWORD on Steemit-English for 100 days. From Figure~\ref{fig-ps-g}, we can see that $F1$ trend of $m$ is that too few or too many words can hurt the F1 of the proposed SEAN-KEYWORD model since fewer words limit the expressive power while more words can include more noises. 
    \item \textbf{\# of document keywords $n$}. The number of document keyword $n$ is set to $\left \{50, 100, 150, 200, 250 \right \}$ for SEAN-KEYWORD. From Figure~\ref{fig-ps-h}, we can see that the influences of $n$ on F1 are limited. The reason might the that when $n > 90$, these keywords contain nearly the entire information of each document. Besides, we can obtain the highest F1 and C\&C at $r=90$, despite the gap is quite small.
\end{itemize}

\begin{figure}[]
    \centering
    \subfigure[Search Depth $L$.]{\includegraphics[width=0.43\textwidth]{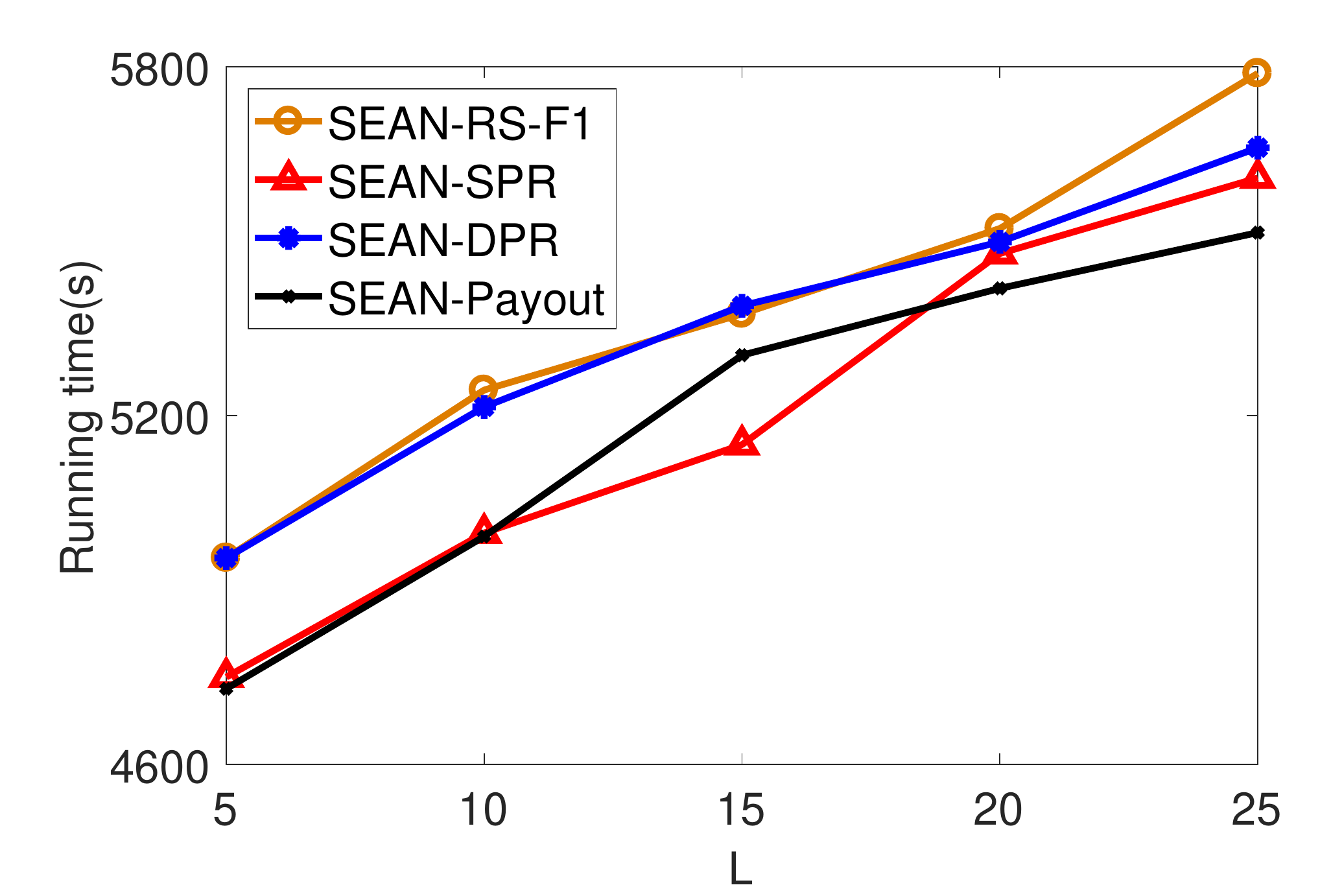}}
    \subfigure[Beam Width $B$.]{\includegraphics[width=0.43\textwidth]{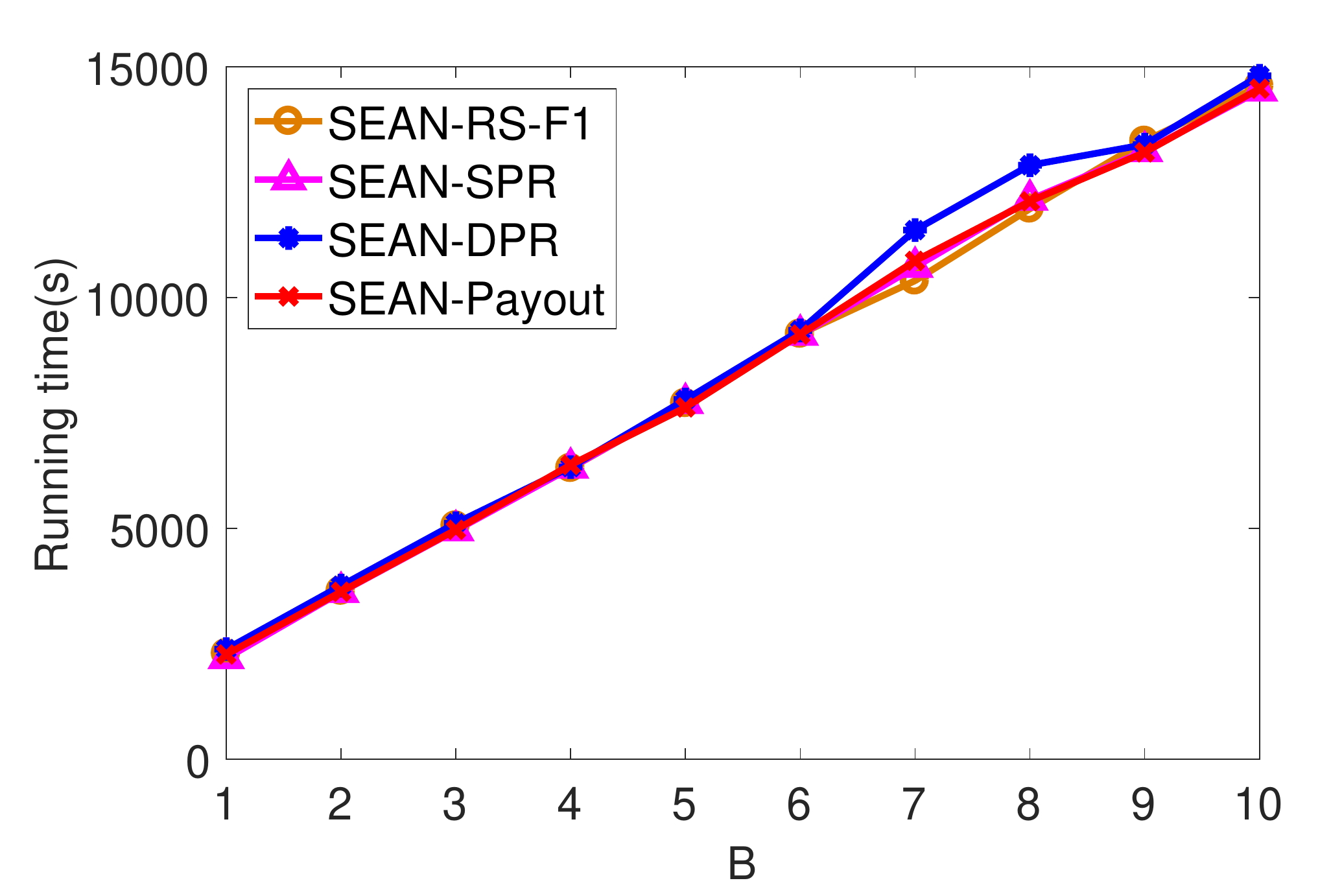}}
    \caption{Running time w.r.t. $L$ and $B$.}
    \label{fig-scalability}
\end{figure}

\subsection{Scalability Analysis}
\label{sec-scalability}

\wenyi{The training complexity of SEAN-END2END and SEAN-KEYWORD are both dominated by the search depth $L$ and the beam width $B$ as explained in Section \ref{sec-complexity}. Since we use the same strategy of social exploration for two models, we only present the scalability analysis on SEAN-END2END for simplicity. }
As shown in Figure~\ref{fig-scalability}, the time consumption of SEAN-END2END with four different exploitation methods is almost linear to $L$ and $ B$.

\section{Related Work}
\label{sec-related}

\subsection{Content Recommendation}
For content recommendation, both content based approaches and CF based approaches have been studied.
Content based approaches use a user's historical reading contents to represent a user, so they are naturally personalized models. 
In~\citep{huang2013learning}, multi-layer neural networks are used to learn embeddings for a given query and related documents.
In~\citep{okura2017embedding}, an end-to-end embedding-based method was proposed to use distributed representations for recommendation.
\citep{wang2018dkn,wang2018ripplenet} apply a knowledge graph to extract latent knowledge-level connections among contents for better exploration.
Moreover, in~\citep{wu2019npa}, the authors propose a neural news recommendation model with
personalized attention.
CF based approaches, including traditional ones~\citep{das2007google,hu2008collaborative,koren2008factorization,koren2009matrix,liu2010personalized,rendle2012factorization,lee2016llorma,zhao2017sloma} and deep learning based models~\citep{wang2015collaborative,covington2016deep,he2017neural,li2017collaborative,sun2018attentive,chen2019social}, are also usually called personalized recommendation, as they consider a user's personal preference based on the user's behaviors or actions on the platform. 




\subsection{Social Recommendation}
In the last decade, social connections have been proved helpful for modeling users' preferences and improving the performance of recommendation. Most of these approaches are based on the social homophily theory \citep{mcpherson2001birds} that people with similar preferences tend to be connected as friends. Therefore, related works to learn users' preferences in RS tend to focus on exploiting social relations to alleviate data sparsity \citep{jiang2012social}. These models could be roughly summarized into the following two categories: classical social recommendation and graph neural networks (GNN) \citep{wu2019neural} in RS. The former ones only use first-order neighbors, most of which perform social regularization in both CF models~\citep{ma2011recommender,wang2013collaborative,hu2015synthetic,zhao2017sloma,zhao2019motif} and content-based models \citep{jiang2012social}. However, social regularization still assumes static social influences between a user and his/her friends. Recent works also apply social attention~\citep{chen2019social,Song0WCZT19,wang2019social,xu2019relation} to calculate different influence strength among neighbors.

GNN-based models applied in RS often perform message passing to aggregate the neighborhood information, such that the $K$-th order social information is captured with $K$ iterations/layers. DiffNet \citep{wu2019neural} proposes to use influence
propagation structure to model the recursive dynamic social
diffusion. GraphRec \citep{fan2019graph} combines user embedding propagation with attention networks to jointly capture interactions and opinions in the user-item graph.
The main difference between our model and others with high-order neighbors  is  that  SEAN  designs  several  new  strategies based on random walk to incorporate high-order friends while GNN-based models have to incorporate all neighbors layer-by-layer, which spend much more computational space. Besides, SEAN considers both consumer satisfaction and creator equality to select the most relative friends while GNN-based approaches only consider the similarity between the target user and his/her neighbors.

\subsection{Exploitation-Exploration}
The exploration-exploitation trade-off is a popular problem which happens in some situations where the model repeatedly learns to make decisions with uncertain pay-offs. This idea is not only highly applied in reinforcement learning \citep{mnih2015human}, but also in other scenarios such as online advertising \citep{li2010exploitation}.

Exploitation-Exploration of items is also a hot topic in RS field~\citep{JoachimsFM97,RadlinskiKJ08,li2010contextual,wang2014exploration,liebman2017designing,zheng2018drn,mcinerney2018explore,chen2019large}.
Exploring more items can introduce more diversity in recommendation results.
However, they still only use the click-through rate (CTR) to evaluate their models.
That means most of them still focus on optimizing the performance of recommendation, which only benefits consumers and the platform.
Moreover, they are still working on traditional user-item based collaborative filtering settings. 
There is a lack of studies on content recommendation and focusing on the creators of the contents.
To our knowledge, we are the first work that considers using the Gini coefficient combined with F1 as a core metric to evaluate different recommendation algorithms.
Some existing work, such as~\citep{SalganikDodds}, has used the Gini coefficient to evaluate Matthew's Effect of a RS. However, they have not simultaneously considered recommendation performance.
In addition to recommendation algorithms, a market-based mechanism design, e.g.,~\citep{WeiMJ05}, is considered as an orthogonal perspective to improve an RS. The related studies to improve the item equality have been shown in~\citep{AbeliukBHHL17,BerbegliaH17}.

\subsection{Attention Mechanisms}
The attention mechanisms have been proved effective in various tasks such as image classification \citep{xu2015show}, language model, machine comprehension \citep{luong2015effective}, and machine translation \citep{bahdanau2014neural}, due to its reasonable assumption that human only focus on selective parts of the whole perception space according to specific tasks. Recently, a hierarchical attention model~\citep{yang2016hierarchical} is proposed to capture patterns of feeds from the word level to the sentence-level and then to the whole documents for the task of document classification. Besides, \citep{velickovic2018graph} presents graph attention networks (GATs) to attend over the friends' features of every node in the citation network. 
In the field of recommendation, \citep{chen2017attentive} introduces both component-level and item-level attention into a CF framework for the multimedia recommendation. \citep{xiao2017attentional} improves factorization machine (FM) by learning the importance of different feature interactions via a neural attention network. KGAT \citep{KGAT19} explicitly models the high-order relations in collaborative knowledge graph by using graph attention networks.

\section{Conclusions and Future Work}
\label{sec-conclusion}
\wenyi{
In this article, we present a social recommendation model, SEAN, which goes beyond personalization by exploring higher-order friends in a social network to help content recommendation.
In the model design and the exploration design, we consider the effects for both content creators and consumers on the social media platform.
This can benefit the platform to attract more innovative content creators and encourage more interactions between the creators and consumers.
We use datasets derived from a decentralized content distribution platform, Steemit, to evaluate our proposed framework.
Experimental results show that we can improve both the creator's equality and consumer's satisfaction in content recommendation.

For future work, we plan to design a more fine-grained approach to select the higher-order friends for users. Specifically, a complete end-to-end model that combines the friends selecting and the recommendation process.
In addition, we notice that our method (and all literature) focuses on modeling single side information in recommendation, i.e., social connections. Therefore, another interesting direction of future work is to investigate how to design an algorithm to well combine various side information to improve recommendation. 
}

\section*{Acknowledgements}
The authors of this paper were supported by NSFC (U20B2053), Hong Kong RGC including Early Career Scheme (ECS, No. 26206717), General Research Fund (GRF, No. 16211520), and Research Impact Fund (RIF, No. R6020-19), and WeBank-HKUST Joint Lab. This article was partially done when the first author was an intern at WeBank AI Department.
We also thank the anonymous reviewers for their valuable comments and suggestions that help improve the quality of this manuscript.


%
%

\bibliographystyle{spbasic}      
\bibliography{ref}

\begin{thebibliography}{65}
\providecommand{\natexlab}[1]{#1}
\providecommand{\url}[1]{{#1}}
\providecommand{\urlprefix}{URL }
\expandafter\ifx\csname urlstyle\endcsname\relax
  \providecommand{\doi}[1]{DOI~\discretionary{}{}{}#1}\else
  \providecommand{\doi}{DOI~\discretionary{}{}{}\begingroup
  \urlstyle{rm}\Url}\fi
\providecommand{\eprint}[2][]{\url{#2}}

\bibitem[{Abeliuk et~al.(2017)Abeliuk, Berbeglia, Hentenryck, Hogg, and
  Lerman}]{AbeliukBHHL17}
Abeliuk A, Berbeglia G, Hentenryck PV, Hogg T, Lerman K (2017) Taming the
  unpredictability of cultural markets with social influence. In: Proceedings
  of the 26th International Conference on World Wide Web

\bibitem[{Bahdanau et~al.(2015)Bahdanau, Cho, and Bengio}]{bahdanau2014neural}
Bahdanau D, Cho K, Bengio Y (2015) Neural machine translation by jointly
  learning to align and translate. In: Proceedings of the 3rd International
  Conference on Learning Representations

\bibitem[{Berbeglia and Hentenryck(2017)}]{BerbegliaH17}
Berbeglia F, Hentenryck PV (2017) Taming the matthew effect in online markets
  with social influence. In: Proceedings of the 31st AAAI Conference on
  Artificial Intelligence

\bibitem[{Chen et~al.(2019{\natexlab{a}})Chen, Zhang, Liu, and
  Ma}]{chen2019social}
Chen C, Zhang M, Liu Y, Ma S (2019{\natexlab{a}}) Social attentional memory
  network: Modeling aspect-and friend-level differences in recommendation. In:
  Proceedings of the 12th ACM International Conference on Web Search and Data
  Mining

\bibitem[{Chen et~al.(2019{\natexlab{b}})Chen, Dai, Cai, Zhang, Wang, Tang,
  Zhang, and Yu}]{chen2019large}
Chen H, Dai X, Cai H, Zhang W, Wang X, Tang R, Zhang Y, Yu Y
  (2019{\natexlab{b}}) Large-scale interactive recommendation with
  tree-structured policy gradient. In: Proceedings of the 33rd AAAI Conference
  on Artificial Intelligence

\bibitem[{Chen et~al.(2017)Chen, Zhang, He, Nie, Liu, and
  Chua}]{chen2017attentive}
Chen J, Zhang H, He X, Nie L, Liu W, Chua TS (2017) Attentive collaborative
  filtering: Multimedia recommendation with item-and component-level attention.
  In: Proceedings of the 40th International ACM SIGIR conference on Research
  and Development in Information Retrieval

\bibitem[{Covington et~al.(2016)Covington, Adams, and
  Sargin}]{covington2016deep}
Covington P, Adams J, Sargin E (2016) Deep neural networks for youtube
  recommendations. In: Proceedings of the 10th ACM Conference on Recommender
  Systems

\bibitem[{Das et~al.(2007)Das, Datar, Garg, and Rajaram}]{das2007google}
Das AS, Datar M, Garg A, Rajaram S (2007) Google news personalization: scalable
  online collaborative filtering. In: Proceedings of the 16th International
  Conference on World Wide Web

\bibitem[{Fan et~al.(2019)Fan, Ma, Li, He, Zhao, Tang, and Yin}]{fan2019graph}
Fan W, Ma Y, Li Q, He Y, Zhao E, Tang J, Yin D (2019) Graph neural networks for
  social recommendation. In: Proceedings of the 28th International Conference
  on World Wide Web

\bibitem[{Grover and Leskovec(2016)}]{grover2016node2vec}
Grover A, Leskovec J (2016) node2vec: Scalable feature learning for networks.
  In: Proceedings of the 22nd ACM SIGKDD International Conference on Knowledge
  Discovery and Data Mining

\bibitem[{He et~al.(2017)He, Liao, Zhang, Nie, Hu, and Chua}]{he2017neural}
He X, Liao L, Zhang H, Nie L, Hu X, Chua TS (2017) Neural collaborative
  filtering. In: Proceedings of the 26th International Conference on World Wide
  Web

\bibitem[{Hu et~al.(2015)Hu, Dai, Song, Huang, and Chen}]{hu2015synthetic}
Hu GN, Dai XY, Song Y, Huang S, Chen J (2015) A synthetic approach for
  recommendation: Combining ratings, social relations, and reviews. In:
  Proceedings of the 24th International Joint Conference on Artificial
  Intelligence

\bibitem[{Hu et~al.(2008)Hu, Koren, and Volinsky}]{hu2008collaborative}
Hu Y, Koren Y, Volinsky C (2008) Collaborative filtering for implicit feedback
  datasets. In: Proceedings of the 8th IEEE International Conference on Data
  Mining

\bibitem[{Huang et~al.(2013)Huang, He, Gao, Deng, Acero, and
  Heck}]{huang2013learning}
Huang PS, He X, Gao J, Deng L, Acero A, Heck L (2013) Learning deep structured
  semantic models for web search using clickthrough data. In: Proceedings of
  the 22nd ACM International Conference on Information \& Knowledge Management

\bibitem[{Jamali and Ester(2010)}]{jamali2010matrix}
Jamali M, Ester M (2010) A matrix factorization technique with trust
  propagation for recommendation in social networks. In: Proceedings of the 4th
  ACM conference on Recommender systems

\bibitem[{Jiang et~al.(2012)Jiang, Cui, Liu, Yang, Wang, Zhu, and
  Yang}]{jiang2012social}
Jiang M, Cui P, Liu R, Yang Q, Wang F, Zhu W, Yang S (2012) Social contextual
  recommendation. In: Proceedings of the 21st ACM international conference on
  Information and knowledge management

\bibitem[{Joachims et~al.(1997)Joachims, Freitag, and Mitchell}]{JoachimsFM97}
Joachims T, Freitag D, Mitchell TM (1997) Web watcher: {A} tour guide for the
  world wide web. In: Proceedings of the 15th International Joint Conference on
  Artificial Intelligence

\bibitem[{Kim(2014)}]{kim2014convolutional}
Kim Y (2014) Convolutional neural networks for sentence classification. In:
  Proceedings of the 2014 Conference on Empirical Methods in Natural Language
  Processing

\bibitem[{Kocsis and Szepesv{\'a}ri(2006)}]{kocsis2006bandit}
Kocsis L, Szepesv{\'a}ri C (2006) Bandit based monte-carlo planning. In:
  European conference on machine learning

\bibitem[{Koehn(2004)}]{koehn2004pharaoh}
Koehn P (2004) Pharaoh: a beam search decoder for phrase-based statistical
  machine translation models. In: Conference of the Association for Machine
  Translation in the Americas

\bibitem[{Koren(2008)}]{koren2008factorization}
Koren Y (2008) Factorization meets the neighborhood: a multifaceted
  collaborative filtering model. In: Proceedings of the 14th ACM SIGKDD
  international conference on Knowledge discovery and data mining

\bibitem[{Koren et~al.(2009)Koren, Bell, and Volinsky}]{koren2009matrix}
Koren Y, Bell R, Volinsky C (2009) Matrix factorization techniques for
  recommender systems. Computer 42(8)

\bibitem[{Lee et~al.(2016)Lee, Kim, Lebanon, Singer, and
  Bengio}]{lee2016llorma}
Lee J, Kim S, Lebanon G, Singer Y, Bengio S (2016) Llorma: local low-rank
  matrix approximation. Journal of Machine Learning Research 17(1):442--465

\bibitem[{Li et~al.(2010{\natexlab{a}})Li, Chu, Langford, and
  Schapire}]{li2010contextual}
Li L, Chu W, Langford J, Schapire RE (2010{\natexlab{a}}) A contextual-bandit
  approach to personalized news article recommendation. In: Proceedings of the
  19th international conference on World Wide Web

\bibitem[{Li et~al.(2010{\natexlab{b}})Li, Wang, Zhang, Cui, Mao, and
  Jin}]{li2010exploitation}
Li W, Wang X, Zhang R, Cui Y, Mao J, Jin R (2010{\natexlab{b}}) Exploitation
  and exploration in a performance based contextual advertising system. In:
  Proceedings of the 16th ACM SIGKDD international conference on Knowledge
  discovery and data mining

\bibitem[{Li and She(2017)}]{li2017collaborative}
Li X, She J (2017) Collaborative variational autoencoder for recommender
  systems. In: Proceedings of the 23rd ACM SIGKDD international conference on
  knowledge discovery and data mining

\bibitem[{Liebman et~al.(2017)Liebman, Khandelwal, Saar-Tsechansky, and
  Stone}]{liebman2017designing}
Liebman E, Khandelwal P, Saar-Tsechansky M, Stone P (2017) Designing better
  playlists with monte carlo tree search. In: Proceedings of the 31st AAAI
  Conference on Artificial Intelligence

\bibitem[{Liu et~al.(2010)Liu, Dolan, and Pedersen}]{liu2010personalized}
Liu J, Dolan P, Pedersen ER (2010) Personalized news recommendation based on
  click behavior. In: Proceedings of the 15th international conference on
  Intelligent user interfaces

\bibitem[{Luong et~al.(2015)Luong, Pham, and Manning}]{luong2015effective}
Luong T, Pham H, Manning CD (2015) Effective approaches to attention-based
  neural machine translation. In: Proceedings of the 2015 Conference on
  Empirical Methods in Natural Language Processing

\bibitem[{Ma et~al.(2011)Ma, Zhou, Liu, Lyu, and King}]{ma2011recommender}
Ma H, Zhou D, Liu C, Lyu MR, King I (2011) Recommender systems with social
  regularization. In: Proceedings of the 4th ACM international conference on
  Web search and data mining

\bibitem[{McInerney et~al.(2018)McInerney, Lacker, Hansen, Higley, Bouchard,
  Gruson, and Mehrotra}]{mcinerney2018explore}
McInerney J, Lacker B, Hansen S, Higley K, Bouchard H, Gruson A, Mehrotra R
  (2018) Explore, exploit, and explain: personalizing explainable
  recommendations with bandits. In: Proceedings of the 12th ACM Conference on
  Recommender Systems

\bibitem[{McPherson et~al.(2001)McPherson, Smith-Lovin, and
  Cook}]{mcpherson2001birds}
McPherson M, Smith-Lovin L, Cook JM (2001) Birds of a feather: Homophily. Annu
  Rev Sociol 27:415--44

\bibitem[{Mnih et~al.(2015)Mnih, Kavukcuoglu, Silver, Rusu, Veness, Bellemare,
  Graves, Riedmiller, Fidjeland, Ostrovski et~al.}]{mnih2015human}
Mnih V, Kavukcuoglu K, Silver D, Rusu AA, Veness J, Bellemare MG, Graves A,
  Riedmiller M, Fidjeland AK, Ostrovski G, et~al. (2015) Human-level control
  through deep reinforcement learning. nature 518(7540):529--533

\bibitem[{Okura et~al.(2017)Okura, Tagami, Ono, and
  Tajima}]{okura2017embedding}
Okura S, Tagami Y, Ono S, Tajima A (2017) Embedding-based news recommendation
  for millions of users. In: Proceedings of the 23rd ACM SIGKDD International
  Conference on Knowledge Discovery and Data Mining

\bibitem[{Van~den Oord et~al.(2013)Van~den Oord, Dieleman, and
  Schrauwen}]{van2013deep}
Van~den Oord A, Dieleman S, Schrauwen B (2013) Deep content-based music
  recommendation. In: Proceedings of the 27th Conference on Neural Information
  Processing Systems

\bibitem[{Radlinski et~al.(2008)Radlinski, Kleinberg, and
  Joachims}]{RadlinskiKJ08}
Radlinski F, Kleinberg R, Joachims T (2008) Learning diverse rankings with
  multi-armed bandits. In: Proceedings of the 25th international conference on
  Machine learning

\bibitem[{Rendle(2012)}]{rendle2012factorization}
Rendle S (2012) Factorization machines with libfm. ACM Transactions on
  Intelligent Systems and Technology 3(3):57

\bibitem[{Rigney(2010)}]{MatthewEffect}
Rigney D (ed)  (2010) The Matthew Effect, How Advantage Begets Further
  Advantage. Columbia University Press

\bibitem[{Salganik et~al.(2006)Salganik, Dodds, and Watts}]{SalganikDodds}
Salganik MJ, Dodds PS, Watts DJ (2006) {Experimental Study of Inequality and
  Unpredictability in an Artificial Cultural Market}. Science
  311(5762):854--856

\bibitem[{Silver et~al.(2016)Silver, Huang, Maddison, Guez, Sifre, van~den
  Driessche, Schrittwieser, Antonoglou, Panneershelvam, Lanctot, Dieleman,
  Grewe, Nham, Kalchbrenner, Sutskever, Lillicrap, Leach, Kavukcuoglu, Graepel,
  and Hassabis}]{MasteringGo2016}
Silver D, Huang A, Maddison CJ, Guez A, Sifre L, van~den Driessche G,
  Schrittwieser J, Antonoglou I, Panneershelvam V, Lanctot M, Dieleman S, Grewe
  D, Nham J, Kalchbrenner N, Sutskever I, Lillicrap T, Leach M, Kavukcuoglu K,
  Graepel T, Hassabis D (2016) Mastering the game of go with deep neural
  networks and tree search. Nature 529:484--503

\bibitem[{Song et~al.(2019)Song, Xiao, Wang, Charlin, Zhang, and
  Tang}]{Song0WCZT19}
Song W, Xiao Z, Wang Y, Charlin L, Zhang M, Tang J (2019) Session-based social
  recommendation via dynamic graph attention networks. In: Proceedings of the
  12th ACM International Conference on Web Search and Data Mining

\bibitem[{Sun et~al.(2018)Sun, Wu, and Wang}]{sun2018attentive}
Sun P, Wu L, Wang M (2018) Attentive recurrent social recommendation. In:
  Proceedings of the 41st International ACM SIGIR Conference on Research \&
  Development in Information Retrieval

\bibitem[{Veli{\v{c}}kovi{\'{c}} et~al.(2018)Veli{\v{c}}kovi{\'{c}}, Cucurull,
  Casanova, Romero, Li{\`{o}}, and Bengio}]{velickovic2018graph}
Veli{\v{c}}kovi{\'{c}} P, Cucurull G, Casanova A, Romero A, Li{\`{o}} P, Bengio
  Y (2018) {Graph Attention Networks}. In: Proceedings of the 6th International
  Conference on Learning Representations

\bibitem[{Wang et~al.(2013)Wang, Chen, and Li}]{wang2013collaborative}
Wang H, Chen B, Li WJ (2013) Collaborative topic regression with social
  regularization for tag recommendation. In: Proceedings of the 23rd
  International Joint Conference on Artificial Intelligence

\bibitem[{Wang et~al.(2015)Wang, Wang, and Yeung}]{wang2015collaborative}
Wang H, Wang N, Yeung DY (2015) Collaborative deep learning for recommender
  systems. In: Proceedings of the 21th ACM SIGKDD International Conference on
  Knowledge Discovery and Data Mining

\bibitem[{Wang et~al.(2018{\natexlab{a}})Wang, Zhang, Wang, Zhao, Li, Xie, and
  Guo}]{wang2018ripplenet}
Wang H, Zhang F, Wang J, Zhao M, Li W, Xie X, Guo M (2018{\natexlab{a}})
  Ripplenet: Propagating user preferences on the knowledge graph for
  recommender systems. In: Proceedings of the 27th ACM International Conference
  on Information and Knowledge Management

\bibitem[{Wang et~al.(2018{\natexlab{b}})Wang, Zhang, Xie, and
  Guo}]{wang2018dkn}
Wang H, Zhang F, Xie X, Guo M (2018{\natexlab{b}}) Dkn: Deep knowledge-aware
  network for news recommendation. In: Proceedings of the 27th International
  Conference on World Wide Web

\bibitem[{Wang et~al.(2014)Wang, Wang, Hsu, and Wang}]{wang2014exploration}
Wang X, Wang Y, Hsu D, Wang Y (2014) Exploration in interactive personalized
  music recommendation: a reinforcement learning approach. ACM Transactions on
  Multimedia Computing, Communications, and Applications 11(1):7

\bibitem[{Wang et~al.(2017)Wang, Yu, Ren, Tao, Zhang, Yu, and
  Wang}]{wang2017dynamic}
Wang X, Yu L, Ren K, Tao G, Zhang W, Yu Y, Wang J (2017) Dynamic attention deep
  model for article recommendation by learning human editors' demonstration.
  In: Proceedings of the 23rd ACM SIGKDD International Conference on Knowledge
  Discovery and Data Mining

\bibitem[{Wang et~al.(2019{\natexlab{a}})Wang, He, Cao, Liu, and Chua}]{KGAT19}
Wang X, He X, Cao Y, Liu M, Chua TS (2019{\natexlab{a}}) Kgat: Knowledge graph
  attention network for recommendation. In: Proceedings of the 25rd ACM SIGKDD
  International Conference on Knowledge Discovery and Data Mining

\bibitem[{Wang et~al.(2019{\natexlab{b}})Wang, Zhu, and Liu}]{wang2019social}
Wang X, Zhu W, Liu C (2019{\natexlab{b}}) Social recommendation with optimal
  limited attention. In: Proceedings of the 25rd ACM SIGKDD International
  Conference on Knowledge Discovery and Data Mining

\bibitem[{Wei et~al.(2005)Wei, Moreau, and Jennings}]{WeiMJ05}
Wei YZ, Moreau L, Jennings NR (2005) A market-based approach to recommender
  systems. {ACM} Trans Inf Syst 23(3):227--266

\bibitem[{Wu et~al.(2019{\natexlab{a}})Wu, Wu, An, Huang, Huang, and
  Xie}]{wu2019npa}
Wu C, Wu F, An M, Huang J, Huang Y, Xie X (2019{\natexlab{a}}) Npa: Neural news
  recommendation with personalized attention. In: Proceedings of the 25rd ACM
  SIGKDD International Conference on Knowledge Discovery and Data Mining

\bibitem[{Wu et~al.(2019{\natexlab{b}})Wu, Sun, Fu, Hong, Wang, and
  Wang}]{wu2019neural}
Wu L, Sun P, Fu Y, Hong R, Wang X, Wang M (2019{\natexlab{b}}) A neural
  influence diffusion model for social recommendation. In: Proceedings of the
  42nd International ACM SIGIR Conference on Research and Development in
  Information Retrieval

\bibitem[{Xiang et~al.(2013)Xiang, Liu, Chen, Xiong, Zheng, and
  Yang}]{xiang2013pagerank}
Xiang B, Liu Q, Chen E, Xiong H, Zheng Y, Yang Y (2013) Pagerank with priors:
  An influence propagation perspective. In: Proceedings of the 33rd
  International Joint Conference on Artificial Intelligence

\bibitem[{Xiao et~al.(2017)Xiao, Ye, He, Zhang, Wu, and
  Chua}]{xiao2017attentional}
Xiao J, Ye H, He X, Zhang H, Wu F, Chua TS (2017) Attentional factorization
  machines: learning the weight of feature interactions via attention networks.
  In: Proceedings of the 37th International Joint Conference on Artificial
  Intelligence

\bibitem[{Xiao et~al.(2019)Xiao, Zhao, Pan, Song, Zheng, and
  Yang}]{xiao2019sean}
Xiao W, Zhao H, Pan H, Song Y, Zheng VW, Yang Q (2019) Beyond personalization:
  Social content recommendation for creator equality and consumer satisfaction.
  In: Proceedings of the 25rd ACM SIGKDD International Conference on Knowledge
  Discovery and Data Mining

\bibitem[{Xu et~al.(2019)Xu, Lian, Han, Li, Xu, and Xie}]{xu2019relation}
Xu F, Lian J, Han Z, Li Y, Xu Y, Xie X (2019) Relation-aware graph
  convolutional networks for agent-initiated social e-commerce recommendation.
  In: Proceedings of the 28th ACM International Conference on Information and
  Knowledge Management

\bibitem[{Xu et~al.(2015)Xu, Ba, Kiros, Cho, Courville, Salakhudinov, Zemel,
  and Bengio}]{xu2015show}
Xu K, Ba J, Kiros R, Cho K, Courville A, Salakhudinov R, Zemel R, Bengio Y
  (2015) Show, attend and tell: Neural image caption generation with visual
  attention. In: Proceedings of the 32nd International Conference on
  International Conference on Machine Learning

\bibitem[{Yang et~al.(2012)Yang, Steck, and Liu}]{yang2012circle}
Yang X, Steck H, Liu Y (2012) Circle-based recommendation in online social
  networks. In: Proceedings of the 18th ACM SIGKDD international conference on
  Knowledge discovery and data mining

\bibitem[{Yang et~al.(2016)Yang, Yang, Dyer, He, Smola, and
  Hovy}]{yang2016hierarchical}
Yang Z, Yang D, Dyer C, He X, Smola A, Hovy E (2016) Hierarchical attention
  networks for document classification. In: Proceedings of the 2016 conference
  of the North American chapter of the association for computational
  linguistics: human language technologies

\bibitem[{Ye et~al.(2012)Ye, Liu, and Lee}]{ye2012exploring}
Ye M, Liu X, Lee WC (2012) Exploring social influence for recommendation: a
  generative model approach. In: Proceedings of the 35th international ACM
  SIGIR conference on Research and development in information retrieval

\bibitem[{Zhao et~al.(2017)Zhao, Yao, Kwok, and Lee}]{zhao2017sloma}
Zhao H, Yao Q, Kwok JT, Lee DL (2017) Collaborative filtering with social local
  models. In: 2017 IEEE International Conference on Data Mining (ICDM)

\bibitem[{Zhao et~al.(2019)Zhao, Zhou, Song, and Lee}]{zhao2019motif}
Zhao H, Zhou Y, Song Y, Lee DL (2019) Motif enhanced recommendation over
  heterogeneous information network. In: Proceedings of the 28th ACM
  International Conference on Information and Knowledge Management

\bibitem[{Zheng et~al.(2018)Zheng, Zhang, Zheng, Xiang, Yuan, Xie, and
  Li}]{zheng2018drn}
Zheng G, Zhang F, Zheng Z, Xiang Y, Yuan NJ, Xie X, Li Z (2018) Drn: A deep
  reinforcement learning framework for news recommendation. In: Proceedings of
  the 27th International Conference on World Wide Web

\end{thebibliography}

\end{document}